%% file: main.tex
\documentclass[twocolumn, secnumarabic,  floatfix, nofootinbib, nobalancelastpage]{revtex4-2}

\def\TitleOfPaper{Conditional probability density functional theory}
\def\preprintlinklocation{http://dft.uci.edu + PAPER REF}

\input Burke_template
\input KieronsMacros

\graphicspath{{figures/}{blue_el/}}

\usepackage[encapsulated]{CJK}

\begin{document}
\begin{CJK*}{UTF8}{gbsn}
\sf
\coloredtitle{\TitleOfPaper}

\coloredauthor{Ryan Pederson}
\email{pedersor@uci.edu}
\affiliation{Department of Physics and Astronomy, University of California, Irvine, CA 92697, USA}

\coloredauthor{Jielun Chen (陈捷伦)}
\affiliation{Department of Physics and Astronomy, University of California, Irvine, CA 92697, USA}

\coloredauthor{Steven R. White}
\affiliation{Department of Physics and Astronomy, University of California, Irvine, CA 92697, USA}

\coloredauthor{Kieron Burke}
\email{kieron@uci.edu}
\affiliation{Department of Chemistry, University of California, Irvine, CA 92697, USA}
\affiliation{Department of Physics and Astronomy, University of California, Irvine, CA 92697, USA}

\date{\today}

\begin{abstract}
We present conditional probability (CP) density functional theory (DFT) as a formally exact theory. In essence, CP-DFT determines the ground-state energy of a system by finding the CP density from a series of independent Kohn-Sham (KS) DFT calculations. By directly calculating CP densities, we bypass the need for an approximate XC energy functional. In this work we discuss and derive several key properties of the CP density and corresponding CP-KS potential. Illustrative examples are used throughout to help guide the reader through the various concepts and theory presented. We explore a suitable CP-DFT approximation and discuss exact conditions, limitations, and results for selected examples.
\end{abstract}
\maketitle
\end{CJK*}


\sec{Introduction}

Over the past 30 years, density functional theory (DFT) has emerged as a widespread tool in many branches of physics and chemistry~\cite{B12}. In particular, Kohn-Sham (KS) DFT is an especially popular method to find ground-state energies and electronic densities of molecules and materials. KS-DFT generally scales better with system size than wavefunction-based methods, such as coupled-cluster theory~\cite{BM07,C66}  or quantum Monte Carlo (QMC)~\cite{AZL12}, and is therefore more suitable for modeling realistic systems. However, KS-DFT relies on finding accurate exchange-correlation (XC) energies, which must be approximated as a functional of spin densities. Hundreds of distinct XC approximations currently exist in standard codes~\cite{LSOM18}, reflecting the tremendous difficulty in finding general and accurate approximations. As these standard KS-DFT approximations usually yield accurate self-consistent densities~\cite{KSB11}, {the resulting errors in energies are primarily due to the flaws in the XC energy approximation itself, i.e. the energy error would not change significantly if evaluated on the exact density.} Moreover, major deficiencies still remain such as the self-interaction error and the inability to derive accurate energies in strongly-correlated systems~\cite{CMW12}.

In the exact theory, one can think of XC energies being determined by the pair density, $P^{\lambda}(\br, \br')$, which is the joint probability density for finding electrons at two points given electron-electron interactions with strength $\lambda / |\br' - \br|$ and fixed ground-state density $\n(\br)$~\cite{LP75}. In standard DFT approximations, the on-top pair densities, $P^{\lambda}(\br, \br)$, are quite accurate~\cite{perdew1997top} and several attempts to use pair densities or density matrices in DFT have been made~\cite{perdew1995escaping,gagliardi2017multiconfiguration,mostafanejad2020global}. This pair density can always be written as
\begin{equation}
P^{\lambda}(\br,\br') = \n(\br)\, \ncprl(\br'),
\label{ncpr}
\end{equation}
where $\ncprl(\br')$ is the conditional probability (CP) density of finding an electron at $\br'$, given an electron at $\br$. 

In standard KS-DFT, the exact KS potential, $v_{\s}[\n](\br)$, is defined to yield $\n(\br)$ in an effective non-interacting electron problem~\cite{DG90}. Analogously, a conditional probability KS potential (CP-KS), $v_{\s}[\ncprl](\br')$, can be defined at each point $\br$ in the system, so that it yields $\ncprl(\br')$ from a KS calculation with $N-1$ electrons. By the Hohenberg-Kohn (HK) theorem~\cite{HK64}, the CP-KS potential, if it exists, is unique. The above equations are for pure density functionals, and their analogs for spin-decomposed systems will be presented in the next sections. This approach, recently presented as conditional probability DFT (CP-DFT) in Ref.~\cite{mccarty2020electronic}, is formally exact. The exact system density and CP-KS potential yield exact pair densities and hence XC energies. However, just like KS-DFT, CP-DFT is only useful if good approximations can be found for a wide array of problems. A clear advantage of CP-DFT is that it naturally has no self-interaction error for one electron. Since standard KS-DFT calculations usually yield highly accurate total system densities~\cite{KSB11}, an accurate CP-KS potential approximation should yield highly accurate XC energies. 

The CP-DFT approach is analogous to the Overhauser model~\cite{overhauser1995pair,davoudi2002self}, which was first proposed to approximate the pair correlation function of the uniform electron gas, but has also been extended to non-uniform systems through the \textit{average pair-density functional theory} (APDFT) approach~\cite{gori2007kohn,gori2005simple}. However, there are key differences: CP-DFT obtains the full pair density whereas APDFT obtains the system- and spherically-averaged pair density, and CP-DFT identifies an exact effective potential within a ground-state KS scheme, the CP-KS potential.

In the context of CP-DFT, authors in Ref.~\cite{mccarty2020electronic} present the \emph{blue electron approximation}. In this rudimentary approximation the CP potential is simply generated by adding $1/|\br' - \br|$ to the external potential. This impurity potential represents the repulsion due to the missing electron as if it were a classical point particle at position $\br$. That is, an electron that is distinguishable from all others (painted blue). The authors use a variant of this approximation to produce surprisingly accurate results. At zero temperature, the correlation energy errors for the uniform gas, simple two electron ions, including H$^-$, and the hydrogen dimer are below $20\%$, even as the bond is stretched. Achieving a smooth accurate binding energy curve for H$_2$ while being in a spin singlet still remains a major challenge to accomplish in KS-DFT. Moreover, as the temperature is increased, the blue electron approximation for the uniform gas becomes more accurate, which is intuitively consistent with the classical limit. 

In this paper, we consider only zero temperature and extend the work of~\cite{mccarty2020electronic} by generalizing to spin-decomposed systems and presenting a formally exact theory for CP-DFT. We give many exact conditions of the CP density and the corresponding CP-KS potential. We illustrate examples using simple 3D and 1D systems. In the latter, we use 1D exponential interactions to mimic the 3D Coulomb interaction~\cite{baker2015one}. We explore the \emph{blue electron approximation} presented in~\cite{mccarty2020electronic} and demonstrate that this approximation satisfies several of these key exact conditions. We discuss failures of the blue electron approximation and suggest improvements.

The paper is structured as follows. In Section~\ref{sec: background} we review DFT and define relevant quantities. In Section~\ref{sec: theory} we present CP-DFT as a formally exact theory. In Section~\ref{sec: exact conditions} we discus additional relevant properties of the CP density and CP-KS potential. In Section~\ref{sec: blue electron approximation} we explore approximations in light of exact conditions. Finally, in Section~\ref{sec: conclusions} we summarize results and discuss future directions of CP-DFT.

\sec{Background and Notation}
\label{sec: background}
\subsection{DFT fundamentals}
We consider non-relativistic purely electronic problems, and use Hartree atomic units (a.u.) throughout the paper. Begin with the variational principal for the exact $N$-electron ground-state energy 
\begin{equation}
     E  = \min_{\Psi} \bra{\Psi} {\hat{H}} \ket{\Psi} \, ,
\end{equation} 
where ${\hat{H}}$ is the $N$-electron Hamiltonian and the search is over all antisymmetric, normalized many-body wavefunctions $\Psi$ ~\cite{L79}. We consider Hamiltonians of the form
\begin{equation}
\hat{H} = \hat{T} + \hat{V}_{\sss ee} + \hat{V} \, ,
\label{eq: many-body hamiltonian}
\end{equation}
where $\hat{T}$ is the usual total kinetic energy operator, $\hat{V}_{\ee}$ is the two-body electron-electron repulsion operator, and $\hat{V}$ is the one-body, possibly spin-dependent, total external potential for the system.

DFT replaces the central role of the one-body spin-dependent potentials, $v(x)$, with the ground-state spin densities $n(x)$, where
\begin{equation}
    n(x) =  N \int dx_2 \cdots dx_N \, | \Psi(x, x_2, \dots, x_N) |^2 \, .
\end{equation} 
Here $x_i = (\br_i, \sigma_i)$ incorporates both spatial and spin coordinates and $\int dx$ is shorthand to denote the integral over all space and sum over both spins and $N$ is the total number of electrons in the system. The total density $n(\br)$ is given by a sum over spin densities, $n(\br) = \sum_{\sigma} n(x)$. From the HK theorem generalized to spin DFT~\cite{BH72, VP75},
there is a one-to-one correspondence between $\{n(x) \}$ and $\{v(x) \}$. From the variational principle, the ground-state energy of a system of $N$-electrons and external potentials $v(x)$ is
\begin{equation}
    E = \min_{\n} \bigg(F[\n] + \int dx \, v(x) \, n(x) \bigg) \, 
\end{equation}
where $F[\n]$ is the universal part of the Hohenberg-Kohn functional, defined as
\begin{equation}
    F[\n] = \min_{\Psi \rightarrow \n} \bra{\Psi} \hat{T} + \hat{V}_{\ee} \ket{\Psi} \, ,
\label{eq: universal functional}
\end{equation}
where the minimization is over all antisymmetric wavefunctions that yield spin-densities $n(x)$~\cite{L79}.

In the KS scheme, there exists an effective one-body potential, $v_{{\sss S}}(x)$, whose corresponding ground-state spin densities for a system of non-interacting electrons match those of the physical interacting system. The total energy of the real system is given in terms of KS quantities:
\begin{equation}
\begin{aligned}
    E = \min_{\n} \bigg(T_{\sss S}[\n] + \int dx \, v(x) \, n(x) 
    + E_{\sss H}[\n] + E_{\xc}[\n] \bigg) \, ,
\end{aligned}
\label{eq: KS minimization}
\end{equation}
where $T_{\sss S}$  is the kinetic energy of the KS non-interacting electrons, $E_{\sss H}$ is the Hartree energy, $n = \n_{\uparrow} + \n_{\downarrow}$ is the total density, and $E_{\xc}$ is the XC energy. We denote the KS wavefunction as $\Phi_{\sss S}[\n]$, which we assume here is a single Slater determinant, as is typical. The KS wavefunction minimizes the kinetic energy functional,
\begin{equation}
    T_{\sss S}[\n] = \min_{\Psi \rightarrow \n} \bra{\Psi} \hat{T} \ket{\Psi} = T[\Phi_{\sss S}[\n]] \, ,
\end{equation}
where
\begin{equation}
    T[\Psi] = - \half \sum_{i=1}^{N} \bra{\Psi} \nabla^2_i \ket{\Psi} \, .
\end{equation}
The Euler-Lagrange equation corresponding to Eq.~\eqref{eq: KS minimization} is 
\begin{equation}
    \frac{\delta T_{\sss S}}{\delta \n(x)} +  v_{{\sss S}}(x) = 0 \, ,
\end{equation}
and all potentials are undetermined up to a constant. The KS spin-dependent potential is
\begin{equation}
    v_{{\sss S}}(x) = v(x) + \intr \frac{n(\br)}{|\br' - \br|} + v_{\xc}[\n](x) \, ,
\end{equation} 
with
\begin{equation}
    v_{{\sss XC}}[\n](x) = \frac{\delta E_{\xc}[\n]}{\delta \n(x)} \, .
\end{equation}
The KS orbitals $\phi_{i \sigma}$ satisfy the KS eigenvalue equation
\begin{equation}
     \bigg[ -\half \nabla^2 + v_{{\sss S}}(x) \bigg] \phi_{i}(x) = \epsilon_{i \sigma} \phi_{i}(x) \, .
\label{eq: KS eig eq}
\end{equation}

\subsection{Adiabatic connection}
\label{sec: adiabatic connection}

In the adiabatic connection approach to KS-DFT~\cite{HJ74, LP75, GL76}, we can modulate the interaction strength $\lambda$ of Coulomb-interacting electrons such that 
\begin{equation}
    F^{\lambda}[\n] = \min_{\Psi^{\lambda} \rightarrow \n} \bra{\Psi^{\lambda}} \hat{T} + \lambda \, \hat{V}_{\ee} \ket{\Psi^{\lambda}} \, .
\label{eq: universal functional lambda}
\end{equation}
For $\lambda = 1$, we have our real, physical system and ground-state density $n$. For $\lambda = 0$, we have the KS system, because we have turned the electron-electron interaction off, but kept the same ground-state density $n$. We can also consider the limit as $\lambda \rightarrow \infty$, which is the strictly correlated electron (SCE) limit where the kinetic energy becomes negligible, see section~\ref{sec: strictly correlated electrons}. In all cases, the ground-state density remains fixed to that of the physical system $n$. This implies that the external potential $v^{\lambda}[n]$ is $\lambda$-dependent. The potential $v^{\lambda}[n]$ corresponds to the unique one-body potential for which $n(\br)$ is the ground-state density for a system of Coulomb-interacting electrons with interaction strength $\lambda$. {As convention, if $\lambda$ is absent from the notation then $\lambda = 1$ is assumed.}  

\subsection{Pair densities and XC holes}

It is natural to discuss the spin-decomposed pair densities $P^{\lambda}(x, x')$ of the $\lambda$-dependent wavefunctions defined in Eq.~\eqref{eq: universal functional lambda}:
\begin{equation}
\begin{aligned}
    &P^{\lambda}(x, x') = \\
    &N(N-1) \int dx_3 \cdots dx_N \, |\Psi^{\lambda}(x, x', x_3, \dots , x_N) |^2 \, ,
\end{aligned}
\label{eq: spin decomp pair densities}
\end{equation}
which is the probability density of finding an electron of spin $\sigma$ at $\br$ and a second electron of spin $\sigma'$ at $\br'$ for interaction strength $\lambda$. The total pair density is the spin-summed quantity:
\begin{equation}
    P^{\lambda}(\br, \br') = \sum_{\sigma \sigma'} P^{\lambda}(x, x') \, ,
\label{eq:pair density}
\end{equation}
which is the joint probability of finding an electron at $\br$ and a second electron at $\br'$ for interaction strength $\lambda$. For $\lambda = 0$, $\Psi^{\lambda = 0} = \Phi_{\sss S}[\n]$ and the pair density is
\begin{equation}
    P^{\lambda = 0}(x, x') = n(x) \, n(x') - |\gamma_s(x, x')|^2\, ,
\end{equation}
where $\gamma_{\sss S}$ is the KS (first-order) density matrix:
\begin{equation}
    \gamma_{\sss S}(x, x') = \delta_{\sigma \sigma'} \sum_{i=1}^{N_{\sigma}} \phi^{*}_{i}(x) \, \phi_{i}(x') \, ,
\end{equation}
The exchange hole is defined as 
\begin{equation}
    n_{\sss X}(x, x') = -\frac{|\gamma_{\sss S}(x, x')|^2}{n(x)} \, ,
\end{equation}
such that 
\begin{equation}
    P^{\lambda = 0}(x, x') = n(x) \big( n(x') + n_{\sss X}(x, x') \big) \, .
\label{eq: pair density lam = 0}
\end{equation}
Checking the normalization allows us to deduce 
\begin{equation}
    \int dx' \, n_{\sss X}(x, x') = -1 \, .
\end{equation}
For $\lambda > 0$, we can generalize Eq.~\eqref{eq: pair density lam = 0} by introducing the $\lambda$-dependent correlation hole $n_{\sss C}^{\lambda}(x, x')$:
\begin{equation}
    P^{\lambda}(x, x') = n(x) \big( n(x') + n_{\sss X}(x, x') + n_{\sss C}^{\lambda}(x, x') \big) \, ,
\end{equation}
where $n_{\sss C}^{\lambda = 0}(x, x') = 0$ and normalization dictates
\begin{equation}
    \int dx' \, n_{\sss C}^{\lambda}(x, x') = 0 \, .
\end{equation}
The XC hole, $n^{\lambda}_{\xc}(x, x')$, is simply the sum of the exchange and correlation holes, $n^{\lambda}_{\xc}(x, x') = n_{\sss X}(x, x') + n_{\sss C}^{\lambda}(x, x')$. Analogously, we can express the spin-summed pair density as
\begin{equation}
    P^{\lambda}(\br, \br') = n(\br) \big( n(\br') + n^{\lambda}_{\xc}(\br, \br') \big) \, ,
\label{eq: spin-summed pair density}
\end{equation}
where
\begin{equation}
    n^{\lambda}_{\xc}(\br, \br') = \bigg( \sum_{\sigma, \sigma'} n(x) \, n^{\lambda}_{\xc}(x, x') \bigg)/n(\br) \, .
\end{equation}
This quantity is especially of interest in DFT since it determines the XC energy in the adiabatic connection integral~\cite{BPE98}:
\begin{equation}
    E_{\xc} = \half\int_0^1 d\lambda \intr \intrp \frac{\n(\br)\, n^{\lambda}_{\xc}(\br, \br')}{|\br-\br'|} \, .
\label{eq: E_xc adiabatic}
\end{equation}

\sec{Theory}
\label{sec: theory}
\subsection{CP-DFT}
\label{subsec: CP-DFT}

\begin{figure*}[ht]
\def\tabularxcolumn#1{m{#1}}
\begin{tabularx}{\linewidth}{@{}cXX@{}}
\begin{tabular}{cc}
\subfloat{\includegraphics[width=0.48\textwidth]{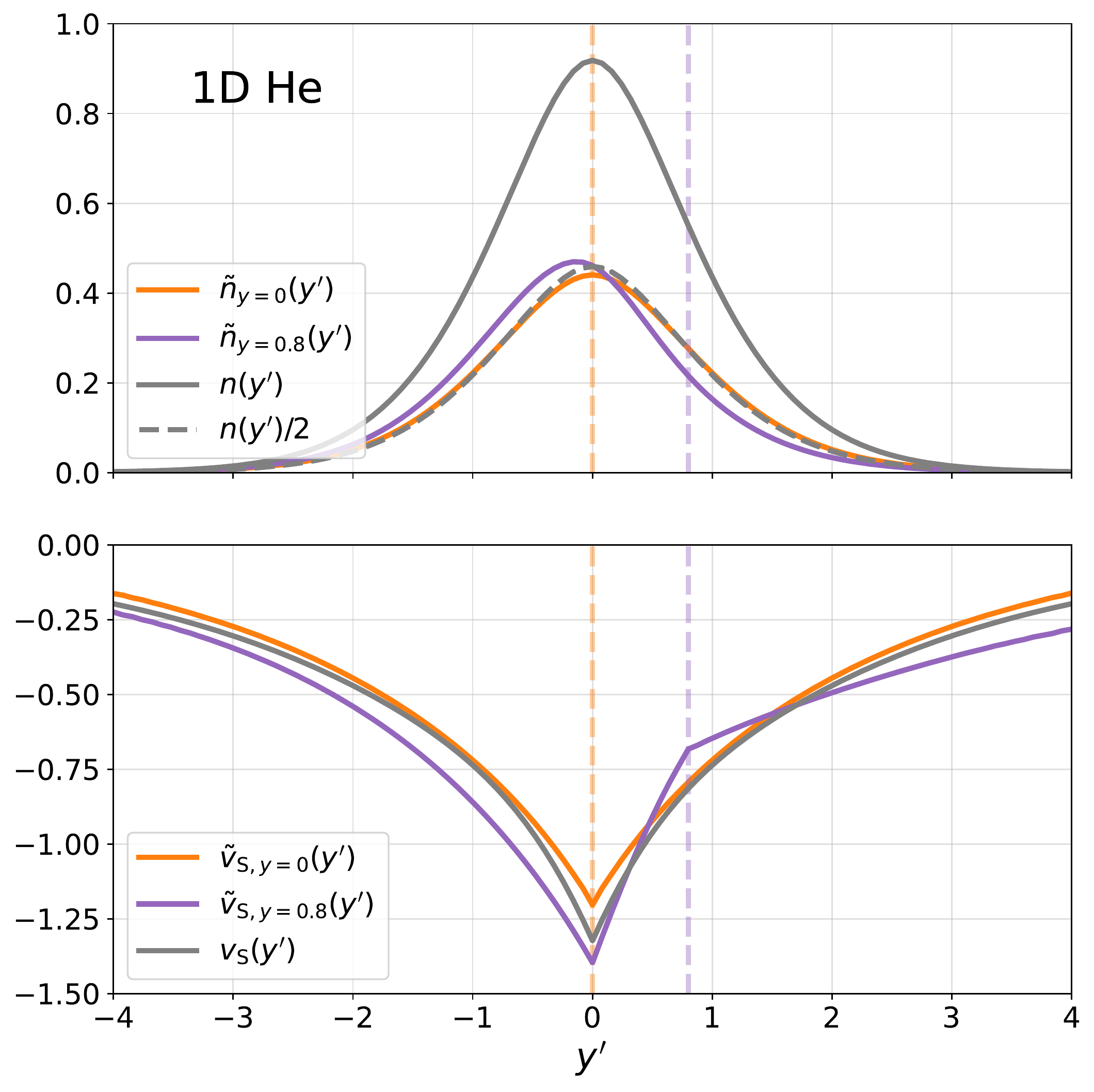}} 
& \subfloat{\includegraphics[width=0.48\textwidth]{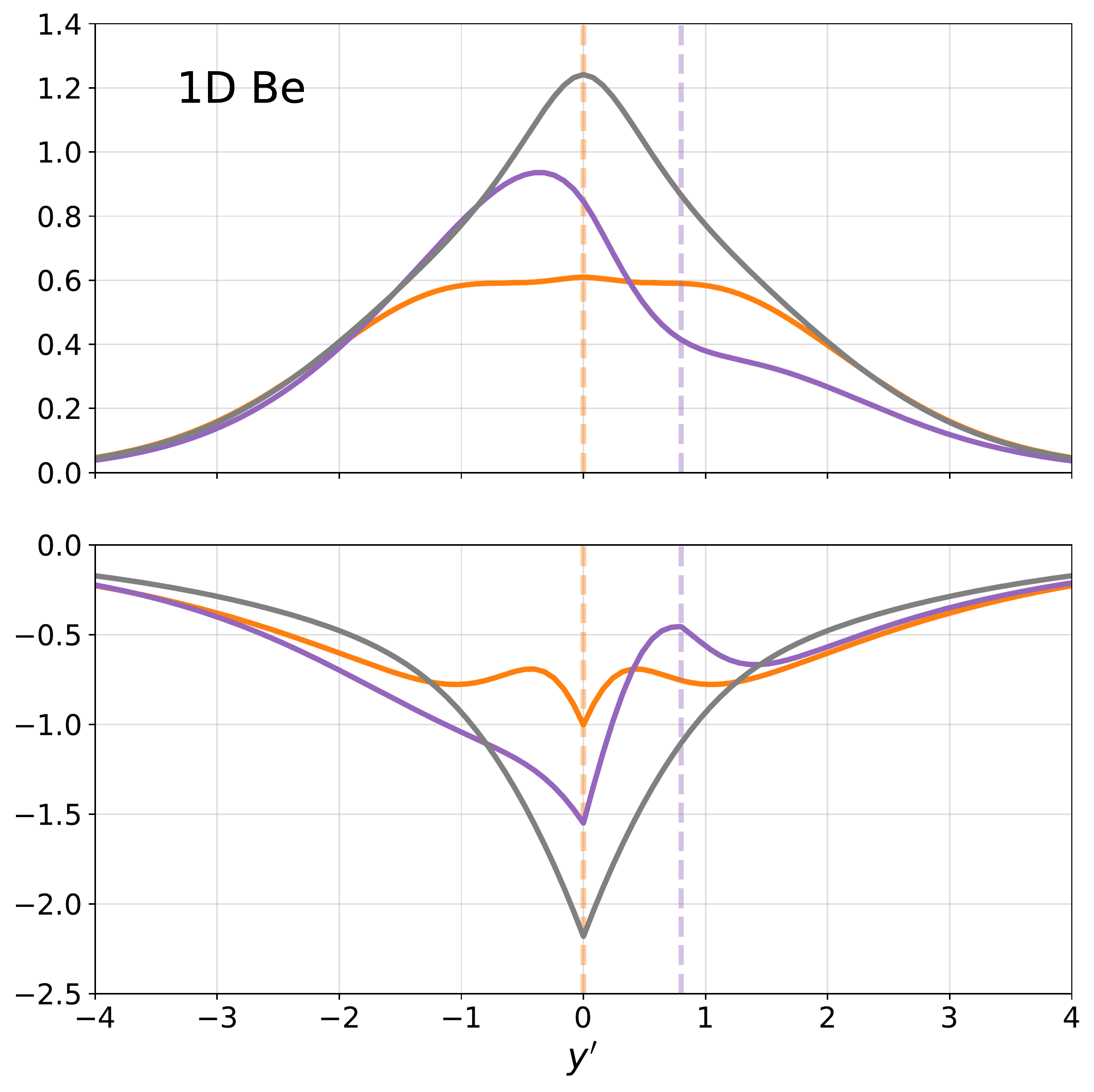}}\\
\end{tabular}
\end{tabularx}

\parbox{\textwidth}{\caption{
Exact CP densities and potentials in 1D He (left) and 1D Be (right): $\tilde v_{{\s}, y}(y')$ is the CP-KS potential with corresponding CP density $\tilde n_y(y')$, $v_{\s}(y')$ the KS potential with corresponding ground-state density $n(y')$. Quantities are plotted for reference positions $y = 0$ and $y = 0.8$. \label{fig: he be example}}}
\end{figure*}

To simplify the discussion, we will first consider the case of $\lambda = 1$ and pure DFT, i.e. not spin decomposed. The subsequent sections will introduce the adaption for different interaction strengths $\lambda$ and spins $\sigma$. 

We define the total conditional probability (CP) density 
density $\tilde\n_{\br}(\br')$ as 
\begin{equation}
    \tilde\n_{\br}(\br') \equiv \frac{P(\br, \br')}{n(\br)}  =  n(\br') + n_{\xc}(\br, \br') \, ,
\end{equation}
which corresponds to the conditional probability density of finding an electron at $\br'$, given an electron at $\br$. The subscript in $\tilde\n_{\br}(\br')$ emphasizes the parametric dependence on the reference position $\br$. In Figure~\ref{fig: he be example} we provide example exact densities for a 1-dimensional (1D) model of the He and Be atoms. 
The 1D model we use throughout mimics reality by replacing 3D Coulomb interactions, $1/ |\br' - \br|$, with 1D exponentials of the form $A \exp(-\kappa |y - y'|)$, where $y, y'\in{\mathbb{R}}$ and $A, \kappa > 0$ are parameters fitted to a soft Coulomb potential, see~\cite{baker2015one, LHPS21, BBW18, LBWB16} for additional information. The purpose of using 1D models here and throughout the paper is to provide simple and illustrative examples which are qualitatively similar to 3D reality. {The 1D ``exact'' density results are obtained from high accuracy density matrix renormalization group (DMRG) calculations using the ITensor library~\cite{itensor} with an energy convergence threshold of $10^{-7}$ Hartree. Associated single-particle potentials were obtained using a modified version of the KS-inversion algorithm outlined in~\cite{FAB18}. The code used to perform KS-inversion is publicly available at ~\cite{dft_1d}.}

In Figure~\ref{fig: he be example} we plot the total ground-state density $n(y')$ for 1D He, which is spatially symmetric and
analogous to a doubly occupied 1s orbital. The exact KS potential $v_{\s}(y')$ corresponding
to this density is also plotted.  We show the CP density $\tilde n_{y}(y')$ for two different reference points,
$y = 0$ and $y = 0.8$. Subtracting the ground-state density $n(y')$ from the CP density yields the ($\lambda=1$)
XC hole for a given reference position. We also show the exchange CP density which, for 2 electrons in a spin singlet, is
simply half the total density.   Thus differences of CP densities from this density are the correlation contributions
which are very tiny when $y=0$.  For $y=0.8$, correlation causes the CP density to be lower in the region of
the electron and greater on the opposite side of the nucleus.

The upper right panel of the figure is the same as the left, but for the Be atom.   
Here the 1D Be ground-state density is analogous to a 1s$^2$2s$^2$ configuration.
Now the CP density integrates to 3 electrons and is not closely approximated by
half the density.  However the same effect occurs when the reference electron is off-center
and charge is pushed to the other side of the nucleus.

The CP density follows several properties. For all reference points $\br$, it must normalize to $N-1$ electrons,
\begin{equation}
    \intrp \tilde\n_{\br}(\br') = N-1 \, . 
\end{equation}
In Figure~\ref{fig: he be example}, taking the area under the curve of any 1D He CP density yields $1$ electron. Similarly, for any 1D Be CP density the integrated area yields $3$ electrons.
Moreover, since the pair density is a symmetrical function of $\br$ and $\br'$, the CP densities satisfy a complementary principle:
\begin{equation}
    \tilde\n_{\br}(\br') = \frac{n(\br')}{\n(\br)}\, \tilde\n_{\br'}(\br),
\label{eq: complementary principle}
\end{equation}
which is simply Bayes' theorem, and may be amenable to modern machine-learning methods~\cite{theodoridis2015machine}. 
In Fig. 1, for Be, this means the ratio of where the purple curve intersects the orange vertical line to where the orange curve 
hits the purple {vertical line equals the ratio of densities (gray curve)} at the two points. For He, since the orange curve is almost identical to the grey dashed line, any CP density (purple curve) passes through (almost) the same value at the origin, namely $n(0)/2$.

Next we turn to the CP potentials that generate CP densities.
For a given reference position~$\br$ and CP density~$\tilde\n_{\br}(\br')$, we define the CP potential, $\tilde v_{\br}(\br')$, as the
unique one-body potential whose total ground-state density for $(N-1)$ Coulomb interacting electrons yields the total CP
density $\tilde\n_{\br}(\br')$. Uniqueness of the CP potential (up to a constant shift) is guaranteed by the HK theorem. 
If it is known, the CP potential for this ($N-1$) electron auxiliary system is simply 
\begin{equation}
    \tilde v_{\br}(\br') = v[\tilde\n_{\br}](\br') 
\end{equation}
where $v[\n](\br)$ is the unique one-body potential for which $\n(\br)$ is the Coulomb-interacting ground-state density. In Figure~\ref{fig: he be example}
the exact CP potentials (and corresponding CP densities) are plotted for He atom at reference points $y=0.0$ and $y=0.8$. Notice that in the $y=0.8$ case, the potential is
asymmetric and contains a kink at the reference position. The kink in the CP potential exists due to the 1D exponential interaction used. 
However, for standard 3D Coulomb interactions, we instead obtain a kink (or cusp) in the CP density at the reference
position due to the electron-electron cusp condition, see section~\ref{sec: cusp condition}.  
A general feature in any dimension is that the CP potential is less negative in the region of the reference point, as it must push electrons away.

We can generalize this HK map to a system of $(N-1)$ electrons with interaction strength $\alpha \geq 0$ and ground-state density $\tilde\n_{\br}$. In analogy, we follow Section~\ref{sec: adiabatic connection}, but throughout we are careful to use $\alpha$ to denote the interaction strength of the associated ($N-1$) electron axiliary system, while $\lambda$ will be used exclusively to denote the interaction strength of the total $N$ electron system with fixed ground-state density $n$. That is, $\alpha$ and $\lambda$ are independent variables in general. For a given $\alpha$, we have
\begin{equation}
    \tilde v_{\br}^{\alpha}(\br') = v^{\alpha}[\tilde\n_{\br}](\br') \, ,
\end{equation}
so that $\tilde v_{{\s}, \br}(\br') = \tilde v_{\br}^{\alpha = 0}(\br')$. 
In Figure~\ref{fig: he be example} exact CP-KS potentials (and corresponding CP densities) are plotted for 1D He and Be atoms. In the
case of the 1D He atom, clearly the CP potential is the CP-KS potential. In the Be atom $y = 0.8$ case, we see a
bump and kink in the CP-KS potential at the reference position, resulting in a small dip in the CP density in the region near the reference position.

The CP-KS potential can be expressed as
\begin{equation}
    \tilde v_{{\s}, \br}(\br') = \tilde v_{\br}^{\alpha}(\br') + \alpha \, v_{\Hx}[\tilde\n_{\br}](\br') + v_{\c}^{\alpha}[\tilde\n_{\br}](\br') \, ,
\label{eq: cp-ks potential w alpha}
\end{equation}
where $v_{\Hx}$ and $v^{\alpha}_{\c}$ are the usual KS-DFT Hartree-exchange and correlation potentials, respectively~\cite{B12}. As in standard KS-DFT, the CP-KS potential can be found self-consistently. Following Eq.~\eqref{eq: KS eig eq}, for each reference point $\br$ the corresponding CP-KS orbitals $\tilde \phi_{\br, i} (\br')$ satisfy the following CP-KS eigenvalue equation
\begin{equation}
     \bigg[ -\half \nabla'^2 + \tilde v_{{\s}, \br}(\br') \bigg] \tilde \phi_{\br, i} (\br') = \epsilon_{\br, i} \, \tilde \phi_{\br, i} (\br') \, ,
\label{eq: CP KS eig eq}
\end{equation}
where $\sum_{i=1}^{N-1} | \tilde \phi_{\br, i} (\br')|^2 = \tilde\n_{\br}(\br')$. 

Notice that by construction, the CP-DFT approach is formally exact: knowledge of an exact CP potential, $\tilde v_{\br}^{\alpha}(\br')$, and a corresponding exact XC potential, $\tilde v_{\xc}^{\alpha}(\br')$, allows a self-consistent KS calculation for an exact CP density $\tilde\n_{\br}(\br')$. 

{By construction, CP densities are nowhere negative and normalized to $N-1$. Thus if their kinetic energy is finite (and we know of no counterexample), they are members of the usual set of well-behaved densities, $\mathcal{I}_N$ of Ref.~\cite{L83}. However, this is insufficient to guarantee non-interacting $v$-representability, just as in standard DFT~\cite{L83,EE83}. In practice, CP-DFT calculations use explicit approximations for both $v_{\xc}$ and $\tilde v_{\br}$, guaranteeing that all CP densities explored are $v$-representable by construction.}

{One may question whether a CP density $\tilde\n_{\br}(\br')$ contains nodes whenever $\br = \br'$ due the exclusion principle, which might make the kinetic energy diverge. However, CP densities are derived from the pair density which is a spin-summed quantity, Eq.~\eqref{eq:pair density}, so we avoid such fermionic nodes. In the spin adaptation of CP-DFT, we must be more careful, see Section~\ref{sec: Spin-adapted CP-DFT}.
}

To extract the XC energy from Eq.~\eqref{eq: E_xc adiabatic} we need the $\lambda$-dependent XC holes, $n^{\lambda}_{\xc}(\br, \br')$, defined in Eq.~\eqref{eq: spin-summed pair density}, so we write
\begin{equation}
    \tilde\n^{\lambda}_{\br}(\br') \equiv \frac{P^{\lambda}(\br, \br')}{n(\br)}  =  n(\br') + n^{\lambda}_{\xc}(\br, \br') \, ,
\end{equation}
for the conditional probability density of finding an electron at $\br'$, given an electron at $\br$ and for interaction strength $\lambda$. The $\lambda$-dependent CP density also normalizes to $N-1$ electrons and satisfies the complementary principle in Eq.~\eqref{eq: complementary principle}.

In analogy with previous formalism, for a given $\br, \lambda, \alpha$ and CP density, $\tilde\n^{\lambda}_{\br}(\br')$, we define the CP potential, $\tilde v_{\br}^{\alpha, \lambda}(\br')$, as the unique one-body potential whose ground-state density for $\alpha$-strength interacting ($N - 1$) electrons yields the CP density, $\tilde\n^{\lambda}_{\br}(\br')$. If it exists, the CP potential is
\begin{equation}
    \tilde v^{\alpha, \lambda}_{\br}(\br') = v^{\alpha}[\tilde\n^{\lambda}_{\br}](\br') 
\end{equation}
where
\begin{equation}
    v^{\alpha}[\tilde\n^{\lambda}_{\br}](\br') = - \frac{\delta F^{\alpha}[\tilde\n^{\lambda}_{\br}] }{\delta \tilde\n^{\lambda}_{\br}} \, .
\end{equation}
The CP-KS potential is $\tilde v^{\lambda}_{{\s}, \br}(\br')$, where $\tilde v^{\lambda}_{{\s}, \br}(\br') = \tilde v^{\alpha = 0, \lambda}_{\br}(\br')$ and
\begin{equation}
    \tilde v^{\lambda}_{{\s}, \br}(\br') = \tilde v_{\br}^{\alpha, \lambda}(\br') + \alpha \, v_{\Hx}[\tilde\n^{\lambda}_{\br}](\br') + v_{\c}^{\alpha}[\tilde\n^{\lambda}_{\br}](\br') \, .
\label{eq: cp-ks broken down}
\end{equation}
Again, we ephasize that, by construction, CP-DFT is formally exact: knowledge of an exact CP potential, if it exists, $ \tilde v_{\br}^{\alpha, \lambda}$, and an exact XC potential, $v_{\xc}^{\alpha}$, allows a self-consistent CP-KS calculation for an exact CP density $\tilde\n^{\lambda}_{\br}$. If such exact CP-KS calculations are carried out for all $\lambda \in [0,1]$ and all reference positions, $\br$, in space, and the exact total ground-state density $n$ is known, then the exact XC energy can be obtained from Eq.~\eqref{eq: E_xc adiabatic}.
  
\subsection{Spin-adapted CP-DFT}
\label{sec: Spin-adapted CP-DFT}

While Section~\ref{subsec: CP-DFT} presents a formally exact construction for \emph{any} interacting electron system, it is useful in practice to have an analogous spin-decomposed formally exact framework. Below we omit $\alpha$ and $\lambda$ dependencies for clarity, but the dependence should follow straightforwardly from the Section~\ref{subsec: CP-DFT}.

We define the spin-CP density $\tilde\n_{x}(\br')$ as
\begin{equation}
    \tilde\n_{x}(\br') \equiv \sum_{\sigma'}  \tilde\n_{x}(x') \, ,
\label{eq: spin-cp density}
\end{equation}
where 
\begin{equation}
    \tilde\n_{x}(x') = \frac{P(x, x')}{n(x)}  =  n(x') + n_{\xc}(x, x') \, .
\end{equation}
The spin-CP density $\tilde\n_{x}(\br')$  has a natural interpretation: given an electron of spin $\sigma$ at reference position $\br$, it is the probability density of finding \emph{any} electron at position $\br'$. The spin-CP density normalizes to $N-1$ electrons,

\begin{equation}
    \intrp \tilde\n_{x}(\br') = N-1 \, ,
\end{equation}
making it a natural object to use as a starting point for spin-adapted CP-DFT. A weighted sum over spin-CP densities can be used to determine the total CP density:
{
\begin{equation}
    \tilde n_{\br}(\br') = \sum_{\sigma} \frac{n_{\sigma}(\br)}{n(\br)} \tilde n_{\br \sigma}(\br') \, .
\label{eq:spin-cp weighted sum}
\end{equation}
}
In the case of a spin-unpolarized system, we recover $\tilde\n_{\uparrow \br}(\br') = \tilde\n_{\downarrow \br}(\br') = \tilde\n_{\br}(\br')$ as expected.

\begin{figure*}[ht]
\def\tabularxcolumn#1{m{#1}}
\begin{tabularx}{\linewidth}{@{}cXX@{}}
\begin{tabular}{cc}
\subfloat{\includegraphics[width=0.48\textwidth]{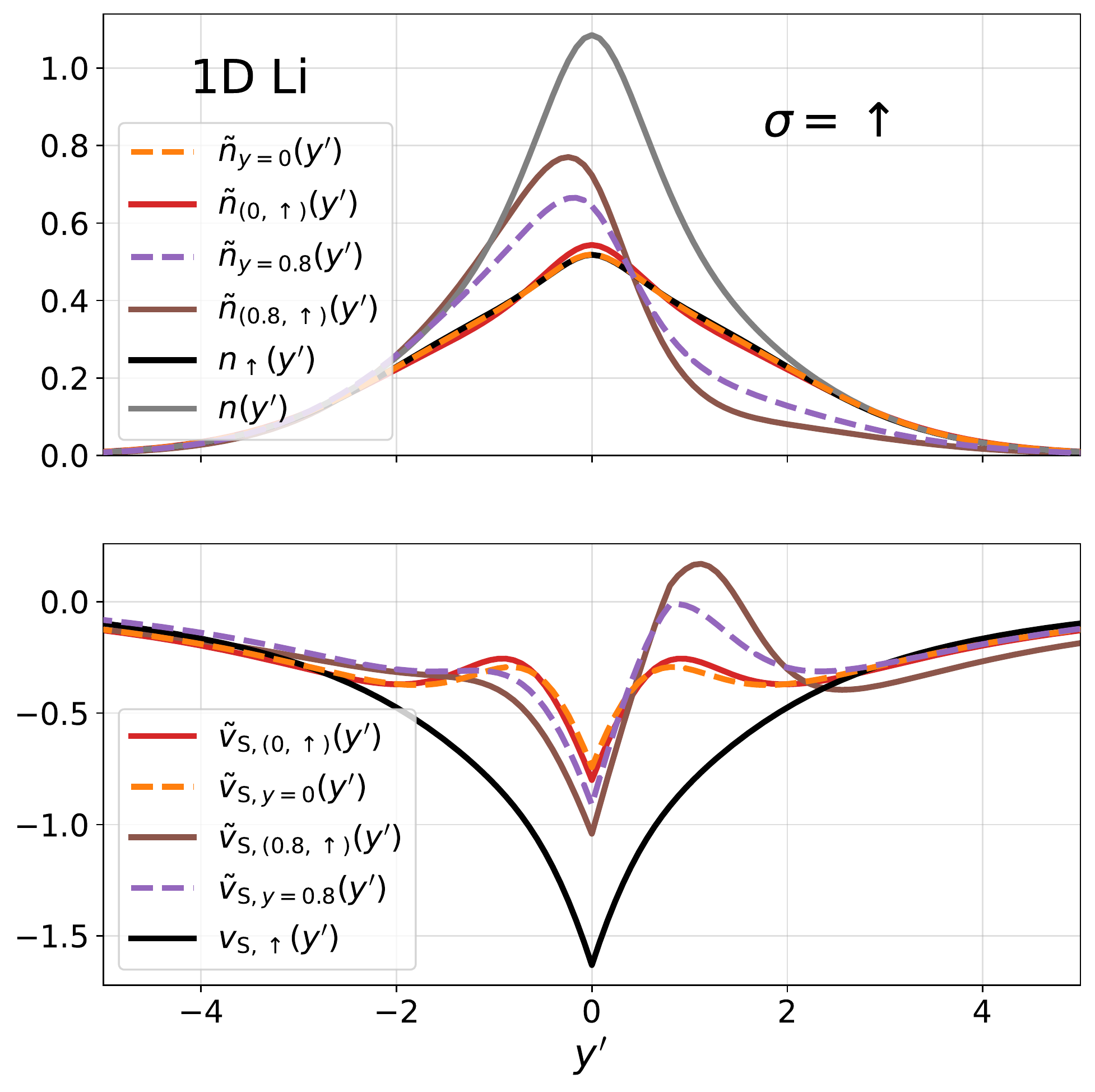}} 
& \subfloat{\includegraphics[width=0.48\textwidth]{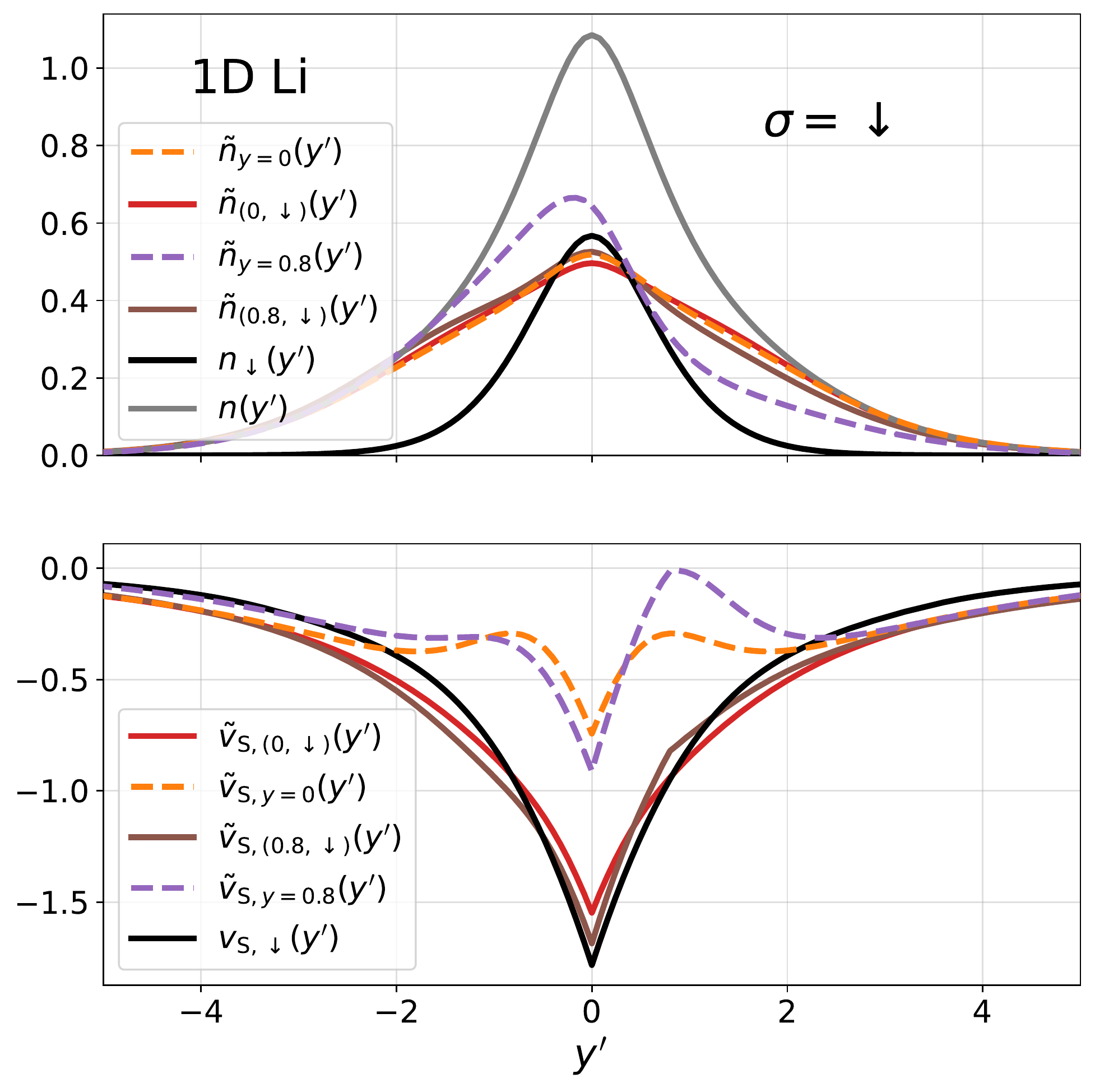}}\\
\end{tabular}
\end{tabularx}

\parbox{\textwidth}{\caption{Exact CP densities and potentials in 1D Li with $N_{\uparrow} = 2, N_{\downarrow} = 1$. {The CP-KS potential is $\tilde v_{{\s}, y}(y')$ with corresponding total CP density $\tilde n_y(y')$ (dashed)}, the spin-CP-KS potential is $\tilde v_{{\s}, x}(y')$ with corresponding spin-CP density $\tilde n_x(y')$, $v_{{\s}, {\sigma}}(y')$ is the KS spin potential with corresponding ground-state spin density $n_{\sigma}(y')$, and $n(y')$ is the total ground-state density. Quantities are plotted for reference space-spin positions $x = (0, \uparrow)$, $x = (0.8, \uparrow)$ (left) and $x = (0, \downarrow)$, $x = (0.8, \downarrow)$ (right).} \label{fig: li example}}

\end{figure*}

In Figure~\ref{fig: li example} we provide exact densities from a 1D model of the spin-polarized Li atom. {For each plotted spin-CP and total CP density, taking the area under the curve yields $2$ electrons. The total CP densities are plotted in Figure~\ref{fig: li example} using dashed curves. From Eq.~\eqref{eq:spin-cp weighted sum}, the total CP density can be obtained from a sum of the spin-CP densities, {with} each weighted by a fraction of density in that spin channel. In 1D Li, for spin-restricted KS orbitals we have the exchange-limit ($\lambda = 0$) relation $n_{\uparrow}{(y')} = \tilde n_{(0, \uparrow)}^{\lambda = 0}{(y')} = \tilde n_{(y, \downarrow)}^{\lambda = 0}{(y')}$. In the upper left panel of Figure~\ref{fig: li example} we see that indeed $n_{\uparrow}{(y')}$ closely approximates the spin-CP density, $\tilde n_{(0, \uparrow)}{(y')}$, with reference point at the origin, which is a high density region {dominated by exchange}. In the upper right panel, for the same reference position but opposite spin, we obtain a very similar spin-CP density, $\tilde n_{(0, \uparrow)}{(y')} \approx \tilde n_{(0, \downarrow)}{(y')}$, {because, for this reference point,} the 1s orbital, which is doubly occupied, is dominant. However, for a spin-down reference point the exchange limit spin-CP density is independent of the reference position, $\tilde n_{(y, \downarrow)}^{\lambda = 0}{(y')} = n_{\uparrow}{(y')}$, which is similar to the case in 1D He (Figure~\ref{fig: he be example}) and correspondingly we see little change in the down spin-CP density, $\tilde n_{(y, \downarrow)}{(y')}$, as the reference position is changed from $0$ to $0.8$. This is not the case for $\sigma = \uparrow$, so we see large differences between $\tilde n_{(0.8, \uparrow)}{(y')}$ and $\tilde n_{(0.8, \downarrow)}{(y')}$.}

The analog definition for the spin-CP potential, $\tilde v_{x}(\br')$, is straightforward: { it is the unique one-body potential whose ``accessible'' ground-state density from ($N-1$) Coulomb-interacting electrons yields the spin-CP density, $\tilde \n_{x}(\br')$, that is, $\tilde v_{x}(\br') = v[\tilde\n_{x}](\br')$. Note that this is within DFT, not spin DFT, and an ``accessible'' ground-state is one that conserves total spin, e.g., the total spins of the $N-$ and $(N-1)$-electron systems cannot differ by more than $1/2$ unit of angular momentum~\cite{EBP96,katriel1980asymptotic}.}

Practically, the main difference from Section~\ref{subsec: CP-DFT} is that, for each reference point $\br$ we solve two independent self-consistent CP-KS equations (one for each spin $\sigma$) to obtain corresponding spin-CP densities $\tilde\n_{x}(\br') = \sum_{i=1}^{N-1} | \tilde \phi_{i, x} (\br')|^2$. { Again, the ground-states should be accessible from a given spin configuration.} For example, in 1D Li with $N_{\uparrow} = 2, N_{\downarrow} = 1$, if the reference electron is spin-down we must singly-occupy the lowest two CP-KS orbitals (with spin-up electrons) in the $(N-1)$ system to conserve spin. If the reference electron is spin-up, we doubly-occupy the lowest CP-KS orbital. This means that similar looking up and down spin-CP densities can have corresponding spin-CP-KS potentials that look quite different. This is why $\tilde v_{{\s}, (0, \uparrow)}{(y')}$ {looks so different from} $\tilde v_{{\s}, (0, \downarrow)}{(y')}$ in Figure~\ref{fig: li example}.

Again, we emphasize that spin-CP densities should be a result from pure DFT with spin-restriction, not spin DFT. One may ask whether a spin-unrestricted CP-KS scheme could be used to obtain the fully spin-decomposed CP spin densities, $\tilde\n_{x}(x')$, which correspond to the conditional probability density of finding an electron at $\br'$ of spin $\sigma'$, given an electron at $\br$ of spin $\sigma$. However, these do not spatially integrate (normalize) to a predetermined integer in general,
\begin{equation}
    \intrp \tilde\n_{x}(x') = N_{\sigma'} - \delta_{\sigma \sigma'} + \intrp n_{\sss C}(x, x') \, ,
\label{eq: integrate fully-spin decomposed densities}
\end{equation}
and so would require an ensemble of fractional particle numbers~\cite{DG90}, presumably making approximations more complicated. Furthermore, these spin densities will contain nodes whenever { $x' = x$ due to the exclusion principal. For instance, in the Li atom (with $N_{\uparrow}=2$ and $N_{\downarrow}=1$), the spin density $\tilde n^{\lambda = 0}_{\br \uparrow}(\br' {\uparrow} )$ is non-$v$-representable, as it is a single-particle density which is not strictly positive~\cite{EE83}.} Therefore, in general we do not recommend fully spin-decomposed CP-DFT. 

\subsection{CP-DFT with averaged quantities}

The spin-CP densities $\tilde\n^{\lambda}_{x}(\br')$ are clearly high-dimensional quantities: in general there is a $\lambda$ and $x$ dependence. In the following we define exact lower-dimensional quantities which can equivalently be used to extract relevant exact energies, such as the electron-electron repulsion potential energy $U_{\ee} = E_{\H} + E_{\xc}$:
\begin{equation}
    U_{\ee} = \half \int_0^1 d\lambda \, \int dx \, n(\br) \intrp \frac{\tilde\n^{\lambda}_{x}(\br')}{|\br' - \br|} \, .
\label{eq: Vee sigma}
\end{equation}
This may be rewritten in terms of purely radial quantities:
\begin{equation}
    U_{\ee} = \frac{N}{2} \sum_\sigma \int du \, u \expval{\tilde n_\sigma (u)} \,    
\end{equation}
where $\bu = \br' - \br$, $u = | \br' - \br|$, and the system and spherically-averaged CP density is
\begin{equation}
    \expval{\tilde n_\sigma (u)} = \frac{1}{N} \int_0^1 d\lambda \, \intr n(\br) \int d\Omega_{\bu} \, \tilde n^{\lambda}_{\sigma \br}(\br + \bu) \, .   
\end{equation}
Note that this radial quantity integrates to $N - 1$. The corresponding radial CP-KS potential is defined as $\tilde v_s[\expval{\tilde n_{\sigma}}](u)$. Knowledge of this much simpler functional also yields the XC energy, but now only a single 1D integral is required, instead of a 3D integral over $\br$. If useful approximations can be found directly for $\tilde v_s[\expval{\tilde n_{\sigma}}](u)$, CP-DFT calculations would be much more efficient. This approach is analogous to \textit{average pair-density functional theory}~\cite{gori2005simple,gori2007kohn} which utilizes an effective radial potential, such as the Overhauser model~\cite{overhauser1995pair,davoudi2002self}, to determine spherically- and system-averaged pair densities.

\subsection{connection to exact factorization?}
\label{subsec: connection to exact factorization}

In this section we present an explicit differential {equation} for $\tilde\n^{\lambda}_{x}(\br')$. We begin by following Ref.~\cite{giarrusso2018response} and partition the $\lambda$-dependent $N$-electron Hamiltonian as
\begin{equation}
    H^\lambda = H_1^\lambda + H_{N-1}^\lambda + \sum_{j=2}^N \frac{\lambda}{|\br-\br_j|} \, ,
\label{eq: lambda hamiltonian}
\end{equation}
where $H_1^\lambda$ is the Hamiltonian for a single electron
\begin{equation}
    H_1^{\lambda} = -\frac{1}{2} \nabla_{\br}^2 + v^{\lambda}[n](\br)
\end{equation}
and $H_{N-1}^\lambda$ is the Hamiltonian for ($N-1$) $\lambda$-strength interacting electrons
\begin{equation}
    H_{N-1}^\lambda = \sum_{i=2}^N \bigg[ -\half \nabla_i^2 + v^{\lambda}[n](\br_i) \bigg] + \half \sum_{i \neq j = 2}^{N} \frac{\lambda}{|\br_i - \br_j|} \, .
\end{equation}
We factorize the normalized ground-state wavefunction $\Psi\l$ of $H^\lambda$ as
\begin{equation}
\Psi\l (x, x_2, \dots, x_N) = {\sqrt{\frac{\n(x)}{N}}}\, \tilde{\Psi}^{\lambda}_{x}(x_2, \dots, x_N) \, .
\label{eq: tildePsi}
\end{equation}
Here, $\tilde{\Psi}^{\lambda}_{x}$ depends parametrically on $x$ and  {is antisymmetric under interchange of the $N-1$ electrons in its argument, but not under interchange of $x$ with any $x_i$, $i \geq 2$, in general.} Note that $\tilde{\Psi}^{\lambda}_{x}$ is not a ground-state wavefunction in general, but is uniquely defined by  Eq. \eqref{eq: tildePsi}, and is often referred to as the~\emph{conditional amplitude}~\cite{giarrusso2018response}. The ground-state wavefunction ${\Psi}^{\lambda}$ yields the spin density $n(x)$, so by construction we have, for all $x$,
\begin{equation}
   \int dx_2 \cdots \int dx_N \, |\tilde{\Psi}^{\lambda}_{x}(x_2, x_3, \dots, x_N) |^2 = 1 \, . 
\end{equation}
We also identify
\begin{equation}
    \tilde\n^{\lambda}_{x}(\br') = (N-1) \sum_{\sigma'} \int dx_3 \cdots dx_N \, |\tilde{\Psi}^{\lambda}_{x}(x', x_3, \dots, x_N)|^2 \, .
\label{eq: cp spin density from wf}
\end{equation}
That is, the total density of $\tilde{\Psi}^{\lambda}_{x}$ is the exact spin-CP density $\tilde\n^{\lambda}_{x}$. Using Eq.~\eqref{eq: tildePsi}, we generalize Ref.~\cite{LPS84} and find an effective equation for $\sqrt{\n(x)}$, 
\begin{equation}
    \bigg[ H_1^{\lambda} + v^{\lambda}_{\text{eff}}(x) \bigg] \sqrt{\n(x)} = E^{\lambda} \sqrt{\n(x)} ,
\label{eq: h_eff}
\end{equation}
where $E^{\lambda}$ is the ground-state energy of $H^\lambda$ and the one-body effective potential is
\begin{equation}
\begin{aligned}
    v^{\lambda}_{\text{eff}}(x) = \lambda \intrp \frac{\tilde\n^{\lambda}_{x}(\br')}{|\br' - \br|} 
    + \bra{\tilde{\Psi}^{\lambda}_{x}} \hat{H}_{N-1}^\lambda \ket{\tilde{\Psi}^{\lambda}_{x}} \\
    + \half \int dx_2 \cdots dx_N \, |\nabla_{\br} \tilde{\Psi}^{\lambda}_{x}|^2 \, .
\end{aligned}
\end{equation}
It is often partitioned this way, where the first term is referred to as the \emph{conditional potential} and the last term is the \emph{kinetic potential}~\cite{giarrusso2018response}. From Eq.~\eqref{eq: h_eff} we also have
\begin{equation}
\begin{aligned}
    \hat{H}^{\lambda}_1 \sqrt{n(x)} \tilde{\Psi}^{\lambda}_{x} = \big(E^{\lambda} - v^{\lambda}_{\text{eff}}(x)\big) \sqrt{n(x)} \tilde{\Psi}^{\lambda}_{x} \\
    - \nabla_{\br} \sqrt{n(x)} \cdot \nabla_{\br} \tilde{\Psi}^{\lambda}_{x}  - \frac{\sqrt{n(x)}}{2} \nabla_{\br}^2 \tilde{\Psi}^{\lambda}_{x} \, .
\end{aligned}
\end{equation}
This yields an apparent Schr\"odinger equation for $\tilde{\Psi}^{\lambda}_{x}$ with Hamiltonian ~\cite{sara_talk_2020}:
\begin{equation}
    \tilde H^{\lambda}_{x} = \hat{H}_{N-1}^\lambda +\sum_{j=2}^N \frac{\lambda}{|\br-\br_j|} + v\l_{\text{nuc}, \, x}(x_2, \dots, x_N)
\label{eq: nuclear hamiltonian}
\end{equation}
where 
\begin{equation}
\begin{aligned}
    &v\l_{\text{nuc}, \, x}(x_2, \dots, x_N) = \\
    &\frac{- 2\nabla_{\br} \sqrt{n(x)} \cdot \nabla_{\br} \tilde{\Psi}^{\lambda}_{x} - \sqrt{n(x)} \nabla_{\br}^2 \tilde{\Psi}^{\lambda}_{x}}{2 \sqrt{n(x)} \tilde{\Psi}^{\lambda}_{x}} 
\end{aligned}
\label{eq: nuclear pot}
\end{equation}
which is a non-multiplicative potential and includes gradients of $\tilde{\Psi}^{\lambda}_{x}$ with respect to $\br$. The full equation reads:
\begin{equation}
    \tilde H^{\lambda}_{x} \tilde{\Psi}^{\lambda}_{x} = \, v^{\lambda}_{\text{eff}}(x) \tilde{\Psi}^{\lambda}_{x} \, .
\label{eq: CP nuc eq}
\end{equation}
This is {\em not} a usual eigenvalue equation that you solve with given boundary conditions~\cite{gonze2018variations}. It is an inhomogenous differential equation satisfied by $\tilde{\Psi}^{\lambda}_{x}$, defined by Eq. \eqref{eq: tildePsi}. The total density of the solution $\tilde{\Psi}^{\lambda}_{x}$ is the exact spin-CP density $\tilde\n^{\lambda}_{x}(\br')$, see Eq.~\eqref{eq: cp spin density from wf}. Eq.~\eqref{eq: CP nuc eq} is an example of the exact factorization technique, which is typically used in studying nuclear dynamics~\cite{AMG10,AMG12,agostini2021ultrafast}, but can also be applied to the pure electronic problem~\cite{SG16,gonze2018variations,kocak2021charge,requist2021fock}. The solution $\tilde{\Psi}^{\lambda}_{x}$ is not always the lowest eigenstate if one treats this as an {inhomogenous eigenvalue problem, see Appendix B in Ref.~\cite{gonze2018variations}. That is, in general, $\tilde{\Psi}^{\lambda}_{x}$ does {\em not} correspond to the $(N-1)$-electron ground-state wavefunction in CP-DFT for CP potential $\tilde v^{\lambda}_{x}(\br') = v[\tilde\n_{x}^{\lambda}](\br')$ and interaction strength $\lambda$. } Worse, $v\l_{\text{nuc}, \, x}$ depends on $N-1$ coordinates simultaneously, so the usual theorems of DFT cannot be applied. For these reasons, the general relationship between CP-DFT and exact electron factorization (EEF) is subtle, contrary to the connections made between the EEF and DFT through Eq.~\eqref{eq: h_eff}~\cite{kocak2022geometric}. However, this analysis can still be useful to CP-DFT in some limits. In Section~\ref{sec: long-range limits}, we discuss the limit $|\br| \to \infty$, where $v\l_{\text{nuc}, \, x}$ vanishes everywhere, and we explore implications to CP-DFT.

\sec{Exact conditions}
\label{sec: exact conditions}

Throughout, it will be convenient to define a CP correction potential, $\Delta \tilde v^{\alpha, \lambda}_{\br}(\br')$, which is simply the difference between the CP potential and the external potential:
\begin{equation}
\begin{aligned}
    \Delta \tilde v^{\alpha, \lambda}_{x}(\br') &\equiv \tilde v^{\alpha, \lambda}_{x}(\br') - v^{\lambda}[n](\br') \\ 
    &= \tilde v^{\alpha}[\tilde n^{\lambda}_{x}](\br') - v^{\lambda}[n](\br') \, .
\end{aligned}
\label{eq: cp correction potential}
\end{equation}

\subsection{Two electron spin-singlet systems}

For $N=1$, $\n\cprl(\br')=0$ since the CP density must normalize to $0$ and thus there is naturally no self-interaction error in CP-DFT for single electron systems~\cite{PZ81}. If $N=2$, the CP density has just one electron and the $\alpha$-dependence and the exact $v_{\Hxc}$ vanish. If the $N = 2$ electron system is unpolarized, i.e. a spin singlet, the CP densities are
\begin{equation}
    \ncprl(\br') = \frac{2 \, |\Psi^{\lambda}(\br,\br')|^2}{\n(\br)}\, ,   
\end{equation}
yielding
\begin{equation}
   \tilde v_{{\s}, \br}^{\lambda}(\br') = \tilde v_{\br}^{\lambda}(\br') =  \half \frac{\nabla'^2 \Psi^{\lambda}(\br,\br')}{\Psi^{\lambda}(\br,\br')} + \epsilon^\lambda_{\br} \, ,
\end{equation}
where $\epsilon^\lambda_{\br}$ is the ground-state enregy of the CP-KS potential, $\tilde v_{{\s}, \br}^{\lambda}(\br')$. Because the wavefunction satisfies the Schr\"odinger equation, we find
\begin{equation}
    \Delta \tilde v_{\br}^{\lambda}(\br') + \Delta \tilde v_{\br'}^{\lambda}(\br) = 
    \frac{\lambda}{|\br' - \br|} - E^{\lambda} + \epsilon^\lambda_{\br'} + \epsilon^\lambda_{\br} \, .
\end{equation}
In the limit $|\br' - \br| \rightarrow 0$ we have to leading order:
\begin{equation}
    \Delta \tilde v_{\br}^{\lambda}(\br') \rightarrow \frac{\lambda}{2|\br' - \br|} \, , \qquad |\br' - \br| \rightarrow 0 \, .
\end{equation}
In the next section, we will see that this exact condition can be more generally derived from the electron-electron cusp condition. In the exchange limit ($\lambda = 0$) for two electron spin-singlet systems we have $ \tilde\n^{\lambda = 0}_{\br}(\br') = \n(\br') / 2$ exactly, which is independent of the reference position. From uniqueness, the corresponding CP-KS potential is then
\begin{equation}
    \tilde v^{\lambda =0}_{{\s}, \br}(\br') = v_{\s}[n](\br') \, .
\label{eq: two el singlet}
\end{equation}
In Figure~\ref{fig: he be example} for 1D He we plot the CP density in the exchange limit (half the total density), which is quite close to the CP densities that include correlation.

\subsection{Cusp condition}
\label{sec: cusp condition}

The generalized electron coalescence cusp condition requires
\begin{equation}
    \frac{\partial  n^{\lambda}_{\xc}(x,u)}{\partial u} \bigg|_{u = 0} = \lambda \, \tilde n^{\lambda}_{x}(\br),
\label{eq: exact n xc cusp condition}
\end{equation}
where $\bu = \br' - \br$, $u = | \br' - \br|$, and the left-hand side has been spherically averaged over $\br+\bu$~\cite{BPE98}. Setting $\lambda = 0$, we see that the cusp condition is a purely correlation-driven effect: the exchange hole does not contribute to the cusp. For generalized Coulomb systems the CP spin density follows 
\begin{equation}
    \frac{\partial \tilde n^{\lambda}_{x}(u)}{\partial u} \bigg|_{u = 0} = \lambda \,  \tilde n^{\lambda}_{x}(\br),    
\label{eq: exact n CP cusp condition}
\end{equation}
for $\br \neq \bf{R}_i$, where $\bf{R}_i$ are nuclei positions. Following Kato's theorem~\cite{kato1957eigenfunctions}, the cusp condition is satisfied in CP-DFT with a CP correction potential that has the following condition: \begin{equation}
    \Delta \tilde v_{\br}^{\alpha, \lambda}(\br') \rightarrow \frac{\lambda}{2|\br' - \br|} \, , \qquad |\br' - \br| \rightarrow 0 \, .
\label{eq: v cp cusp}
\end{equation}

\subsection{Long-range limits}
\label{sec: long-range limits}

When the reference position $\br$ is sent to infinity, we can generalize Ref.~\cite{EBP96} for arbitary $\lambda$ to deduce the leading-order term of the ground-state wavefunction of the $\lambda$-interacting $N$-electron system as
\begin{equation}
    \lim_{|\br| \rightarrow \infty} \Psi\l (x, x_2, \dots, x_N) = {\sqrt{\frac{\n(x)}{N}}}\, \tilde{\Psi}^{\lambda, N-1}_{\hat{x}}(x_2, \dots, x_N) \, ,
\label{eq: large r psi tilde}
\end{equation}
where $\tilde{\Psi}^{\lambda, N-1}_{\hat{x}}$ is {an accessible} ground-state of the ionized $(N-1)$-system and is parametrically dependent on $\hat{x} = (\hat{\br}, \sigma)$, where $\hat{\br} = \br/|\br|$ is the direction and $\sigma$ is the spin of the electron sent to infinity. The $\hat{\br}$ dependence occurs when the ionized ($N-1$)-system ground-state is otherwise degenerate~\cite{EBP96}. The wavefunction $\tilde{\Psi}^{\lambda, N-1}_{\hat{x}}$ is a solution to Eq.~\eqref{eq: CP nuc eq} at the large $\br$ limit. For a finite number of degenerate $(N-1)$ ground-states we expect $\tilde{\Psi}^{\lambda, N-1}_{\hat{x}}$ to remain fixed for small changes in $\hat{\br}$, therefore the gradients in Eq.~\eqref{eq: nuclear pot} vanish and $v^{\lambda}_{\text{nuc}, \, x}$ vanishes (see Refs.~\cite{gori2016asymptotic} and ~\cite{gori2018asymptotic} for an in-depth discussion and exceptions). Taking $\alpha = \lambda$, we identify to leading order:  
\begin{equation}
    \Delta \tilde v^{\lambda = \alpha}_{x}[\n](\br') \rightarrow \frac{\lambda}{|\br' - \br|} \, , \qquad |\br| \rightarrow \infty \, .
\label{eq: large r correction pot}
\end{equation}
That is, in the large reference position limit, the CP correction potential in Eq.~\eqref{eq: large r correction pot} simply represents an impurity perturbation that breaks the possible degeneracy in the ionized $(N-1)$-system ground-state.

We continue this analysis and denote the fully spin-decomposed density of the wavefunction $\tilde{\Psi}^{\lambda, N-1}_{\hat{x}}$ as $\tilde{\n}^{\lambda, N-1}_{\hat{x}}(x')$ and the total density of this wavefunction as $\tilde{\n}^{\lambda, N-1}_{\hat{x}}(\br')$. The latter corresponds to the asymptotic limit of the spin-CP density to leading order:
\begin{equation}
    \lim_{|\br| \rightarrow \infty} \tilde{\n}^{\lambda}_{x}(\br') = \tilde{\n}^{\lambda, N-1}_{\hat{x}}(\br') \, .
\end{equation}
From the complementary principal, for spin-CP densities we obtain:
\begin{equation}
    \lim_{|\br'| \rightarrow \infty} \tilde{\n}^{\lambda}_{x}(\br') = \frac{\sum_{\sigma'} n(x') \, \tilde{\n}^{\lambda, N-1}_{\hat{x}'}(x)}{n(x)} \, .
\end{equation}
For unpolarized systems or the total CP density we obtain
\begin{equation}
    \lim_{|\br'| \rightarrow \infty} \frac{\tilde{\n}^{\lambda}_{\br}(\br')}{n(\br')} =  \frac{\tilde{\n}^{\lambda, N-1}_{\hat{\br}'}(\br)}{n(\br)} \, .
\label{eq: large rp CP density}
\end{equation}
That is, for a given reference position $\br$, the asymptotic behavior of the total CP density is the same as the asymptotic behavior of the total density of the system, up to a multiplicative constant. {This can be seen in Figure~\ref{fig: he be example}. For Be, the ionized $N-1$ system has a 1s orbital {very similar to that of} the neutral system, so for $y = 0$ the ratio appearing in Eq.~\eqref{eq: large rp CP density} is roughly unity and the CP density and total density match {far from the nucleus}. For He and $y = 0$ the ratio is roughly $1/2$ and we see that the CP density and half the total density overlap closely {far away}.}

At large $|\br'|$ the highest occupied KS orbital dominates the density ~\cite{LPS84} and we have
\begin{equation}
    \bigg[ -\half \nabla'^2 + v_{\s}[n](\br') \bigg] \sqrt{\n(\br')} = \epsilon_{\text{HOMO}} \sqrt{\n(\br')} \, .
\label{eq: KS eqs HOMO}
\end{equation}
Since $\tilde{\n}^{\lambda}_{\br}(\br')$ is simply proportional to $n(\br')$ at large $|\br'|$, from uniqueness (up to a constant shift) we have
\begin{equation}
     \tilde v_{{\s}, \br}^{\lambda}(\br') \rightarrow v_{\s}[n](\br')\, , \qquad |\br'| \rightarrow \infty \, .
\label{eq: large rp v_s cp}
\end{equation}
The CP-KS potential approaches the original KS potential far from the system. This can also be seen in Figure~\ref{fig: he be example}, where the CP-KS potentials approach the original KS potential {far from the nucleus}. We can use Eq.~\eqref{eq: large rp v_s cp} to determine a corresponding exact condition for the CP correction potential {defined in Eq.~\eqref{eq: cp correction potential}.} From Eqs.~\eqref{eq: cp-ks potential w alpha} and~\eqref{eq: cp correction potential} we have
\begin{equation}
\begin{aligned}
    \tilde v_{{\s}, \br}^{\lambda}(\br') = v^{\lambda}[n](\br') + \Delta \tilde v^{\alpha, \lambda}_{\br}(\br') + v_{\Hxc}^{\alpha}[\tilde n^{\lambda}_{\br}](\br') \\
    = v_s[n](\br') - v_{\Hxc}^{\lambda}[n](\br') + \Delta \tilde v^{\alpha, \lambda}_{\br}(\br') + v_{\Hxc}^{\alpha}[\tilde n^{\lambda}_{\br}](\br') \, .
\end{aligned}
\label{eq: cp-ks potential lambda and alpha}
\end{equation}
The asymptotic limit $|\br'| \rightarrow \infty$ of the usual KS-DFT potentials are well-known to leading order~\cite{ECMV92}:
\begin{equation}
    v_{\Hxc}^{\lambda}[n](\br') \rightarrow \frac{\lambda (N-1)}{r'} \, , \qquad |\br'| \rightarrow \infty \, .
\label{eq: v_hxc long range}
\end{equation}
Since the CP density $\tilde n^{\lambda}_{\br}$ integrates to $N-1$ electrons:
\begin{equation}
    v_{\Hxc}^{\alpha}[\tilde n^{\lambda}_{\br}](\br') \rightarrow \frac{\alpha (N-2)}{r'} \, , \qquad |\br'| \rightarrow \infty \, .
\label{eq: large rp v_hxc}
\end{equation}
From Eqs.~\eqref{eq: large rp v_s cp}~-~\eqref{eq: large rp v_hxc} we obtain
\begin{equation}
    \Delta \tilde v^{\alpha, \lambda}_{\br}(\br') \rightarrow \frac{N(\lambda-\alpha) + 2\alpha - \lambda}{r'}, \qquad |\br'| \rightarrow \infty \, .
\label{eq: large rp}
\end{equation}

\subsection{Strictly correlated electron limit}
\label{sec: strictly correlated electrons}

When $\lambda \to \infty$, we approach the strictly correlated electron (SCE) limit. The Hamiltonian of Eq.~\eqref{eq: lambda hamiltonian} has the expansion~\cite{giarrusso2018response}
\begin{equation}
    H^{\lambda} \rightarrow \lambda \big( \hat V_{\ee} + \hat V^{\sce}\big) + \mathcal{O}(\sqrt{\lambda}), \qquad \lambda \to \infty \, ,
\label{eq: H sce}
\end{equation}
where the last term is kinetic and is subleading. In this limit we have defined 
\begin{equation}
    \hat V^{\lambda}[n] \rightarrow  \lambda \hat V^{\sce}[n] = \lambda \sum_{i = 1}^N v^{\sce}[n](\br_i), \qquad \lambda \to \infty \, ,
\end{equation}
where $\hat V^{\sce}[n]$ is the one-body potential that minimizes the classical potential potential energy operator $\hat V_{\ee} + \hat V^{\sce}[n]$ and delivers $\n(\br)$ as the ground-state density. The ground-state wavefunction of such a Hamiltonian collapses into a distribution that can be expressed as~\cite{mirtschink2012energy}
\begin{equation}
\begin{aligned}
    &|\Psi^{\sce} (x, x_2, \dots, x_N)|^2 = \\
    &\frac{1}{N!} \sum_{\mathcal{P}} \int d^3s \, \frac{\n(\bf s)}{N}
    \prod_{i=1}^{N} \delta^{(3)}(\br_i - {\bf f}_{\mathcal{P}(i)}({\bf s})) \, ,
\end{aligned}
\end{equation}
where $\mathcal{P}$ denotes a permutation of $1, \dots, N$, ensuring $|\Psi^{\sce}|^2$ is symmetric with respect to exchanging the coordinates of identical particles. The co-motion functions, ${\bf f}_i({\br})$, dictate the positions of correlated electrons given an electron at position $\br$. The co-motion functions satisfy cyclic group properties, with ${\bf f}_1({\br}) \equiv \br$, ${\bf f}_2({\br}) \equiv {\bf f}(\br)$, ${\bf f}_3({\br}) = {\bf f}({\bf f}(\br))$, and so on such that ${\bf f}_{N+1}({\br}) = \br$. See Ref.~\cite{seidl2017strictly} for additional properties. In the SCE limit, the pair density becomes
\begin{equation}
\begin{aligned}
    P^{\sce}(\br, \br') = 
    \sum_{i \neq j = 1} \int d^3s \, \frac{\n({\bf s})}{N}
    \delta^{(3)}(\br - {\bf f}_{i}({\bf s})) \, \delta^{(3)}(\br' - {\bf f}_{j}({\bf s})) \\
    = \sum_{i \neq j = 1} \int d^3s \, \frac{\n({\bf s})}{N}
    \delta^{(3)}({\bf f}_{N - i + 2}(\br) - {\bf s}) \, \delta^{(3)}({\bf f}_{N - j + 2}(\br') - {\bf s}) .
\end{aligned}
\end{equation}
We can neglect terms where neither $i$ nor $j$ are equal to $1$ since the delta functions in the integral will not overlap for differing non-trivial co-motion functions. Evaluating the integral and applying cyclic properties we obtain
\begin{equation}
    P^{\sce}(\br, \br') = \n(\br) \sum_{i = 2}^{N} \delta^{(3)}(\br' - {\bf f}_i(\br)) \, .
\end{equation}
The CP density is then
\begin{equation}
    \tilde n_x^{\sce}(\br') = \sum_{i = 2}^{N} \delta^{(3)}(\br' - {\bf f}_i(\br)) \, .
\end{equation}
Partitioning the SCE Hamiltonian in Eq.~\eqref{eq: H sce} like Eq.~\eqref{eq: lambda hamiltonian} and neglecting the subleading gradient terms allows us to identify a corresponding CP correction potential for $\alpha, \lambda \rightarrow \infty$,
\begin{equation}
    \lambda \, \Delta \tilde v^{\sce}_{x}[\n](\br') = \frac{\lambda}{|\br' - \br|} \, .
\label{eq: sce correction pot}
\end{equation}
This is the potential required in this semiclassical limit. In this limit, we have simply fixed the ``missing'' electron at position $\br$ as if it were a distinguishable particle and solve the resulting ($N - 1$) electron system in the presence of this impurity potential, Eq.~\eqref{eq: sce correction pot}. In the context of classical statistical mechanics, this is equivalent to the Percus test particle procedure~\cite{P62,C91,ACE17}.

\begin{figure*}[ht]
\def\tabularxcolumn#1{m{#1}}
\begin{tabularx}{\linewidth}{@{}cXX@{}}
\begin{tabular}{cc}
\subfloat{\includegraphics[width=0.48\textwidth]{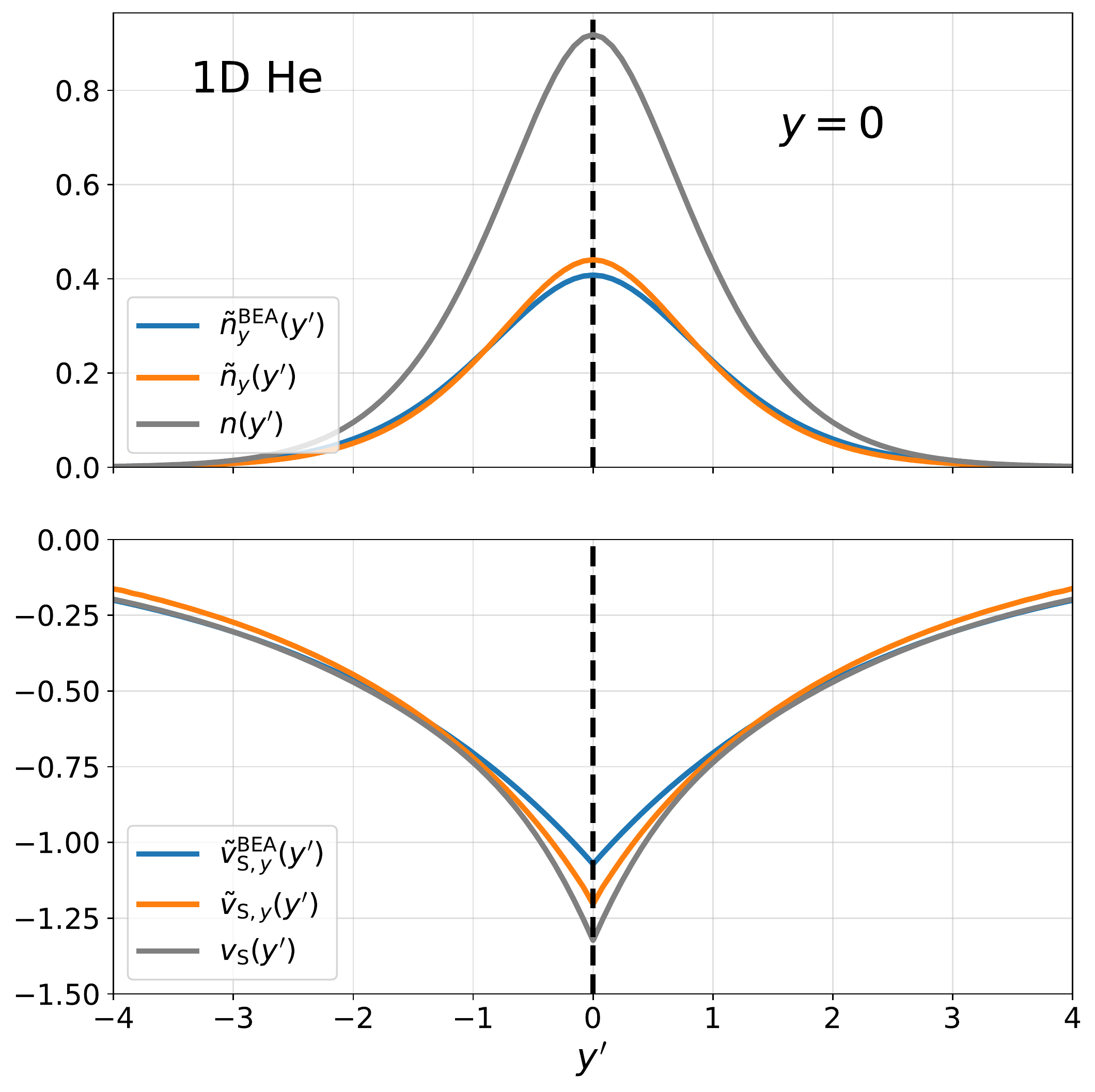}} 
& \subfloat{\includegraphics[width=0.48\textwidth]{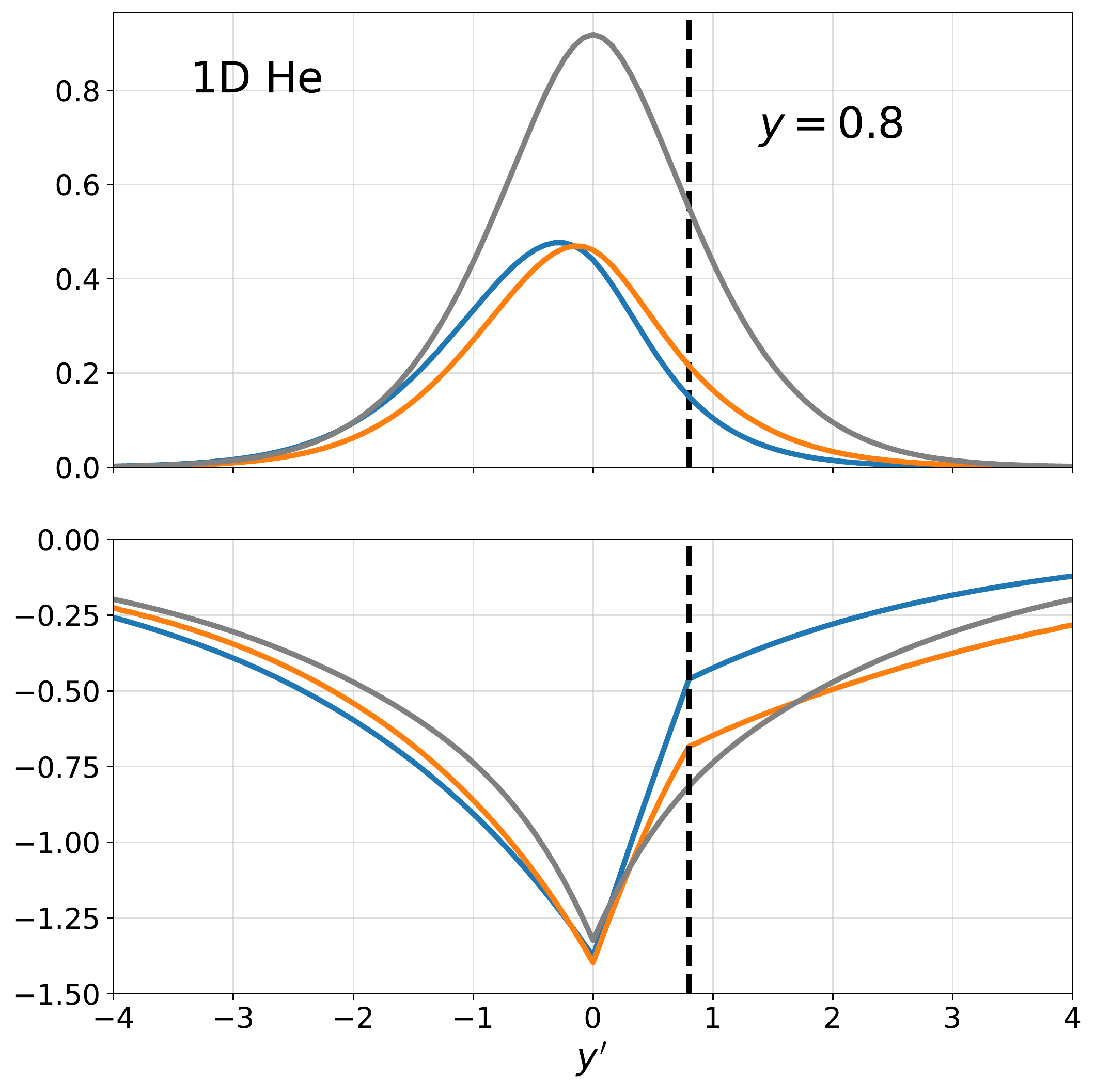}}\\
\end{tabular}
\end{tabularx}

\parbox{\textwidth}{\caption{
CP densities and potentials in 1D He ($\lambda = 1$): $\tilde v_{{\s}, y}^{\mathrm{BEA}}(y')$ is the 1D BEA approximation to the CP-KS potential with corresponding CP density $\tilde n_y^{\mathrm{BEA}}(y')$, $\tilde v_{{\s}, y}(y')$ is the exact CP-KS potential with corresponding exact CP density $\tilde n_y(y')$, $v_{\s}(y')$ the exact KS potential with corresponding exact ground-state density $n(y')$, and $v(y')$ the external potential for our 1D He atom. Quantities are plotted for reference position $y = 0$ (left) and for $y = 0.8$ (right). \label{fig: he example with bea}}}

\end{figure*}

\subsection{Dissociation limit}

Consider a simple example of bond dissociation, the stretched H$_2$ bond. At the dissociation limit, the exact wavefunction has the Heitler-London~\cite{HL27} form ($\lambda \neq 0$):
\begin{equation}
\Psi\l(\br_1,\br_2)=\frac{1}{\sqrt{2}}\left( \phi_A(\br_1)\, \phi_B(\br_2) + \phi_B (\br_1)\, \phi_A(\br_2) \right)
\label{eq: H2 HL}
\end{equation}
where $\phi_A$ and $\phi_B$ are atomic H orbitals localized on each of the two  protons. This yields a CP density:
\begin{equation}
\tilde \n^{\lambda}_{\br}(\br')=\n_B(\br'),~~~~~\br~{\rm near}~A \, ,
\end{equation}
and vice versa, and vanishes elsewhere. Thus the total electron-electron repulsion potential energy from Eq.~\eqref{eq: Vee sigma} vanishes due to the lack of overlap, and each atomic region correctly yields the one-electron energy of a isolated hydrogen atom.

We can also consider arbitrarily long {neutral} H$_N$ chains. In the spin-singlet case, where $N$ is an even number, we obtain a generalized version of Eq.~\eqref{eq: H2 HL} at the dissociation limit ($\lambda \neq 0$):
\begin{equation}
    \Psi^{\lambda}(\br_1, \dots, \br_N) = \frac{1}{\sqrt{N!}} \sum_{\mathcal{P}} 
    \prod_{i=1}^{N} \, \phi_i(\br_{\mathcal{P}(i)}) \, ,
\end{equation}
where $\phi_i$ are atomic H orbitals localized on nuclei $i$, with position ${\bf R}_i$, in the enumerated H$_N$ chain. The CP density is:
\begin{equation}
\tilde \n^{\lambda}_{\br}(\br')= \sum_{i \neq j = 1}^N \n_i(\br'),~~~~~\br~{\rm near}~{\bf R}_j \, .
\end{equation}
Again, the total electron-electron repulsion potential energy from Eq.~\eqref{eq: Vee sigma} vanishes due to the lack of overlap, and each atomic region correctly yields a one-electron energy of a isolated hydrogen atom. The corresponding CP potential is
\begin{equation}
    \tilde v^{\lambda, \alpha}_{\br}(\br') = - \sum_{i \neq j = 1}^N \frac{1}{|\br' - {\bf R}_i|} ,~~~~~\br~{\rm near}~{\bf R}_j \, .
\label{eq: cp n h chains}
\end{equation}
Hence the CP correction potential is then
\begin{equation}
    \Delta \tilde v^{\lambda, \alpha}_{\br}(\br') = \frac{1}{|\br' - {\bf R}_j|} ,~~~~~\br~{\rm near}~{\bf R}_j \, .
\end{equation}

\begin{figure*}[ht]
\def\tabularxcolumn#1{m{#1}}
\begin{tabularx}{\linewidth}{@{}cXX@{}}
\begin{tabular}{cc}
\subfloat{\includegraphics[width=0.48\textwidth]{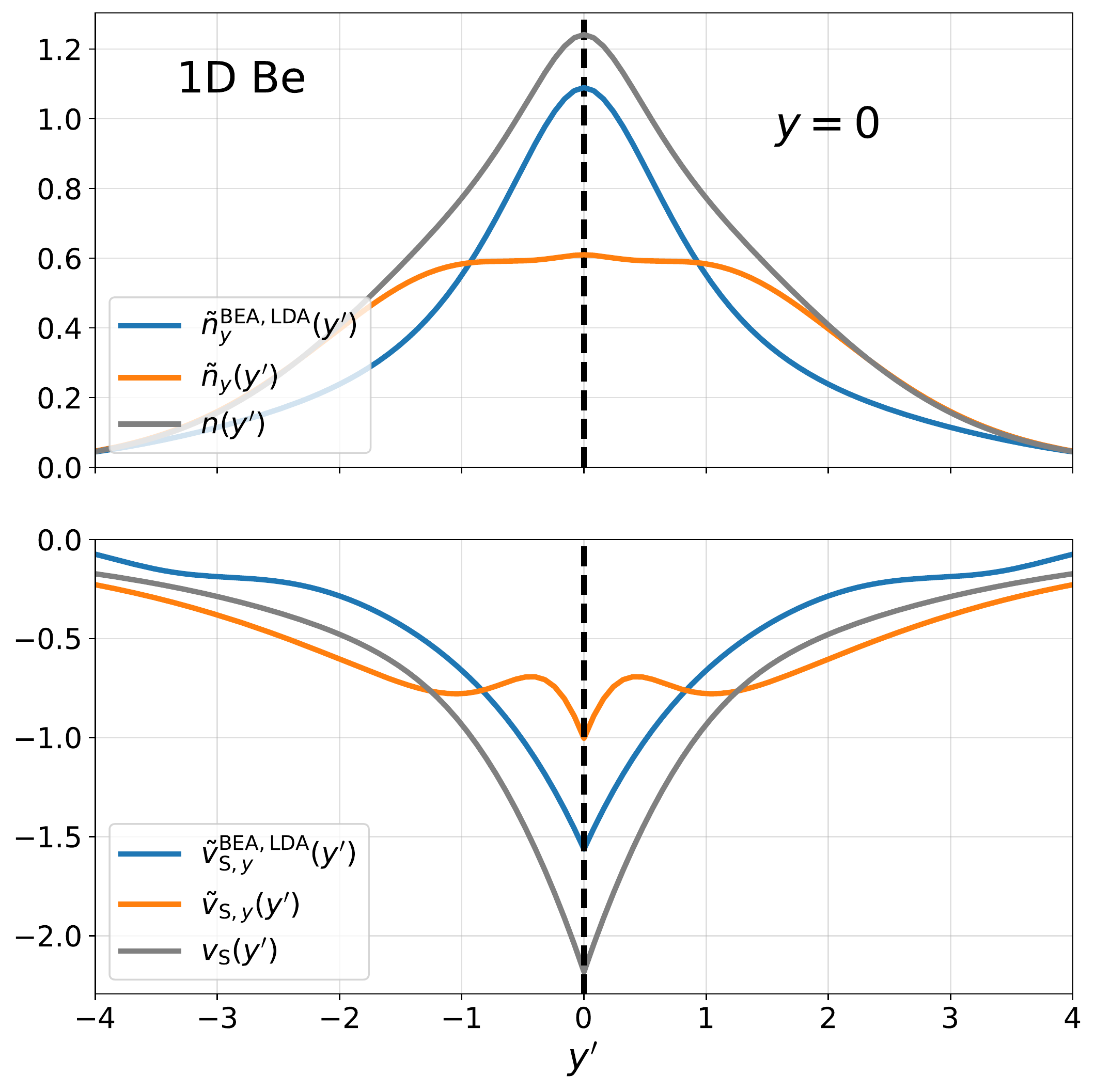}} 
& \subfloat{\includegraphics[width=0.48\textwidth]{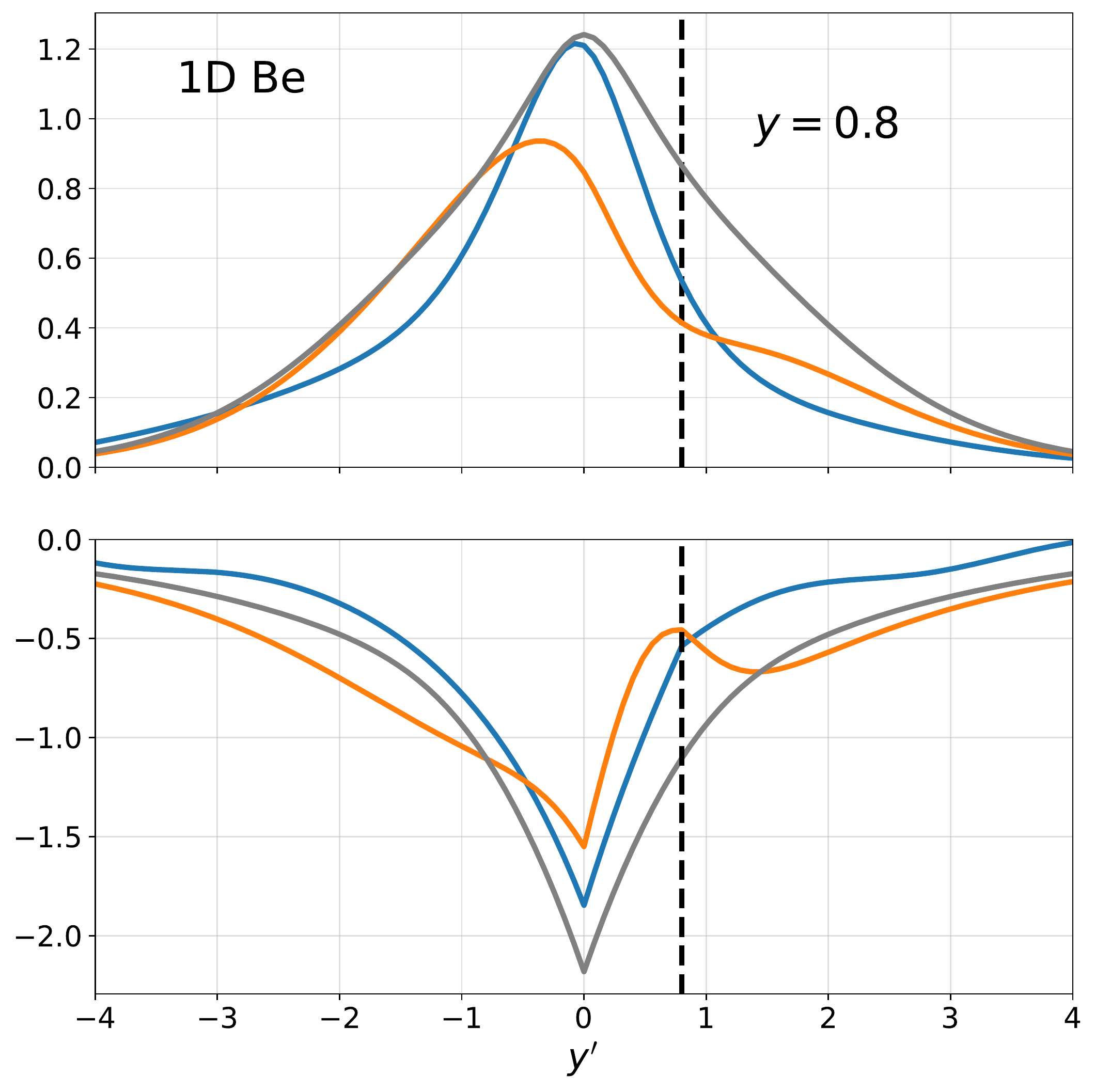}}\\
\end{tabular}
\end{tabularx}

\parbox{\textwidth}{\caption{CP densities and potentials in 1D Be ($\lambda = 1$): $\tilde v_{{\s}, y}^{\mathrm{BEA, LDA}}(y')$ is the 1D BEA approximation to the CP-KS potential using the 1D LDA~\cite{baker2015one} approximation, see Eq.~\eqref{eq: blue el cp-ks potential w lda}. The corresponding CP density is $\tilde v_{{\s}, y}^{\mathrm{BEA, LDA}}(y')$. \label{fig: be example with bea}}}

\end{figure*}

\sec{Blue electron approximation}
\label{sec: blue electron approximation}

In the \emph{blue electron approximation} (BEA) we make the simple approximation
\begin{equation}
\Delta \tilde v^{\lambda = \alpha, \, \mathrm{BEA}}_{\br}(\br') \equiv \frac{\lambda}{|\br'-\br|}
\label{eq: blue el approx}
\end{equation}
to the CP correction potential defined in Eq.~\eqref{eq: cp correction potential}. 
We call this the blue electron approximation because it corresponds to the classical result, Eq.~\eqref{eq: sce correction pot}. It is as if our reference electron was a distinguishable particle, painted blue for instance, fixed at position $\br$, yielding an impurity potential, Eq.~\eqref{eq: blue el approx}.
Using the BEA and setting $\lambda = \alpha$ in Eq.~\eqref{eq: cp-ks potential lambda and alpha} we obtain
\begin{equation}
    \tilde v_{{\s}, \br}^{\lambda, \mathrm{BEA}}(\br') = v_s[n](\br') + \frac{\lambda}{|\br'-\br|} + v_{\Hxc}^{\lambda}[\tilde n^{\lambda}_{\br}](\br') - v_{\Hxc}^{\lambda}[n](\br') \, .
\label{eq: blue el cp-ks potential}
\end{equation}
While simple and semiclassical in origin, the BEA satisfies a surprising number of exact conditions discussed in this work. 

{In the exchange limit ($\lambda = 0$), the BEA CP correction potential and $v_{\Hxc}^{\lambda = 0}$ terms vanish, leaving $\tilde v_{{\s}, \br}^{\lambda=0, \mathrm{BEA}}(\br')~=~v_s[n](\br')$. For two electron spin-singlet systems, this is exact, see Eq.~\eqref{eq: two el singlet}, {but not otherwise}. Therefore, in general we do not expect the BEA to perform well for $N > 2$, unless exchange-limit corrections are also incorporated into the CP correction potential.}

Taking a reference position $\br$ that is far away from the system we see that BEA matches the exact asymptotic behavior in Eq.~\eqref{eq: large r correction pot} by construction. Similarly, taking $\lambda = \alpha$ in Eq.~\eqref{eq: large rp}, we obtain 
\begin{equation}
    \Delta \tilde v^{\lambda = \alpha}_{\br}(\br') \rightarrow \frac{\lambda}{r'}, \qquad |\br'| \rightarrow \infty \, ,
\end{equation}
{which matches the asymptotics of BEA in Eq.~\eqref{eq: blue el cp-ks potential} if the exact asymptotic decay of $v_{\Hxc}^{\lambda}{(\br')}$, Eq.\eqref{eq: v_hxc long range}, is assumed. However, for approximate XC functionals, especially local and semilocal functionals, this condition is {usually} violated~\cite{LB94}.}

We can also show that BEA correctly dissociates {neutral} H$_N$ chains. {In the dissociation limit, the exact $v_{\Hxc}^{\lambda}{(\br')}$ terms will vanish and we obtain:}
\begin{equation}
    \tilde v_{{\s}, \br}^{\lambda, \mathrm{BEA}}(\br') = \frac{\lambda}{|\br'-\br|} - \sum_{i=1}^N \frac{1}{|\br' - {\bf R}_i|} \, .
\end{equation}
In the absence of the first term, this KS potential would yield an $N$-fold degenerate ground-state eigenvalue which is equal to the H atom ground-state energy. For $\lambda \neq 0$ and when $\br$ is near $\{ {\bf R}_i \}$, the first term breaks some degeneracy: the ground-state eigenvalue has the same value as before but is now $(N-1)$-fold degenerate. The associated ground-state CP-KS orbital, $\tilde \phi_{0, \br}$, is
\begin{equation}
    \tilde \phi_{0, \br}(\br') = \frac{1}{\sqrt{N-1}} \sum_{i \neq j = 1}^N \phi_i(\br')  \, ,~~~~~\br~{\rm near}~{\bf R}_j \, ,
\label{eq: h dissoc cp-ks orbitals}
\end{equation}
where $\phi_i$ are atomic H orbitals centered at position ${\bf R}_i$. The resulting CP density is the exact one, Eq.~\eqref{eq: cp n h chains}. {If an approximate $v_{\Hxc}^{\lambda}{(\br')}$ is used, and there exists a self-interaction error for $1$ electron systems, the atomic H orbitals in Eq.~\eqref{eq: h dissoc cp-ks orbitals} are correspondingly approximate.}

\begin{figure*}[ht]
\def\tabularxcolumn#1{m{#1}}
\begin{tabularx}{\linewidth}{@{}cXX@{}}
\begin{tabular}{cc}
\subfloat{\includegraphics[width=0.48\textwidth]{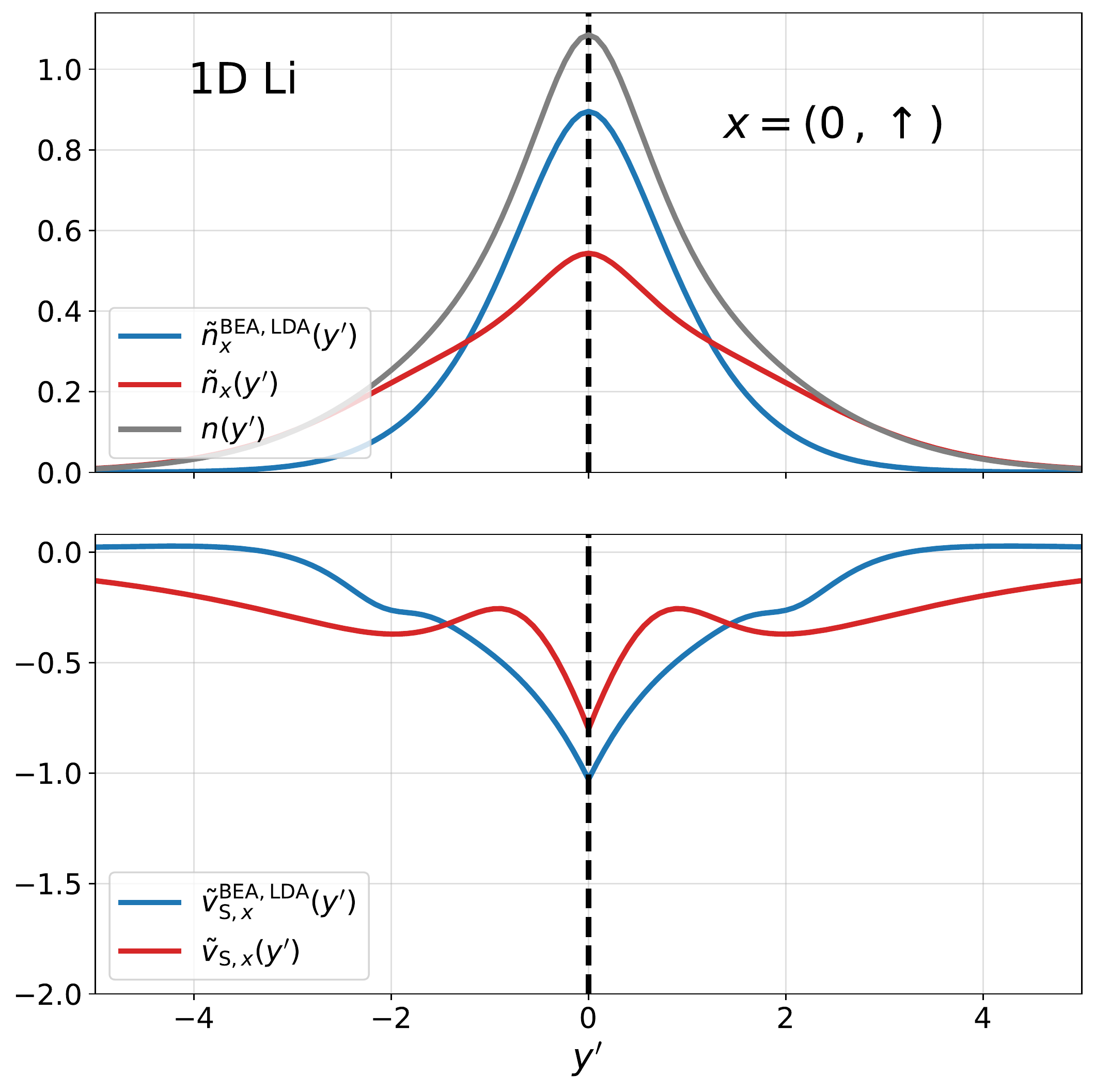}} 
& \subfloat{\includegraphics[width=0.48\textwidth]{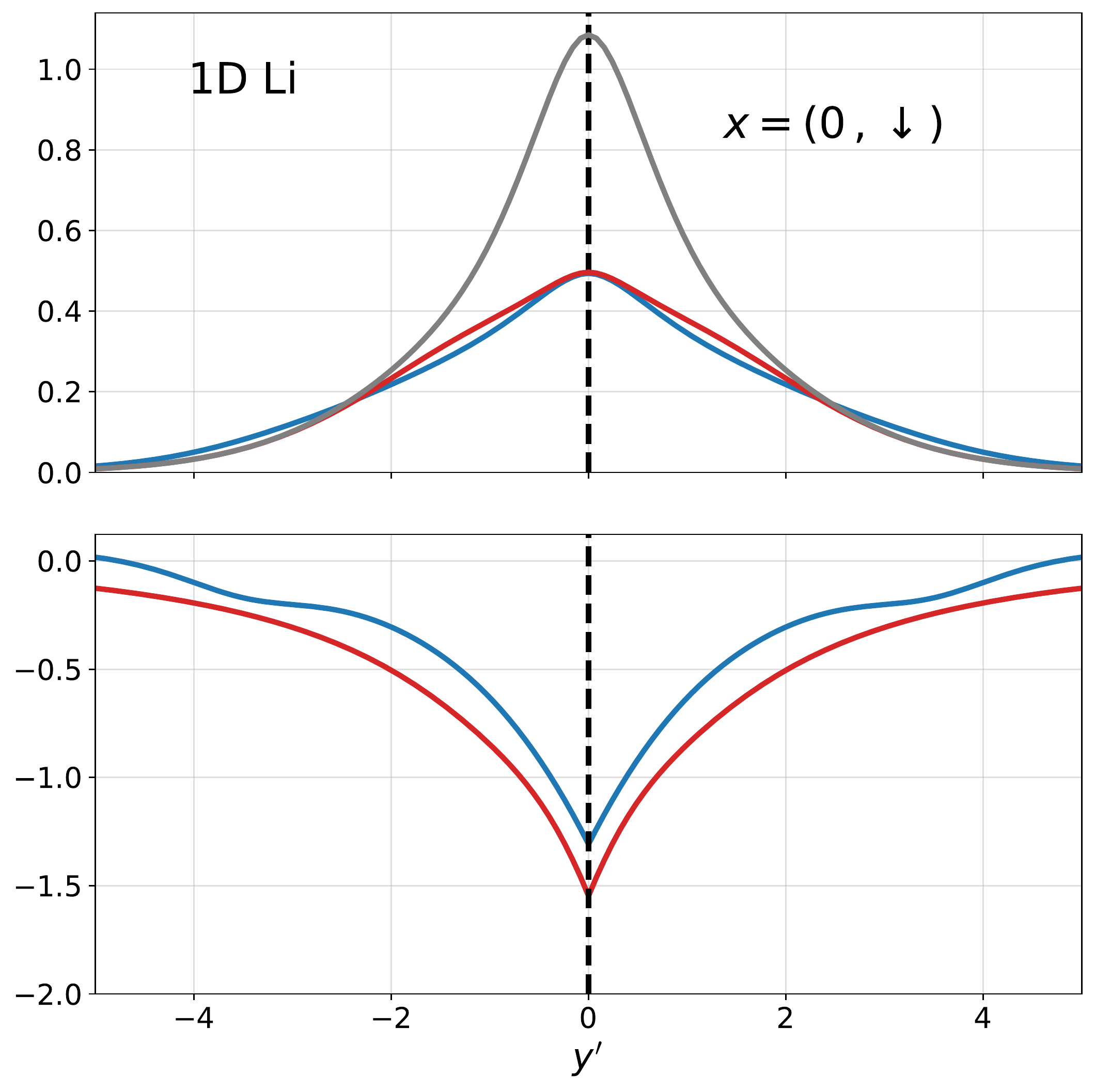}}\\
\subfloat{\includegraphics[width=0.48\textwidth]{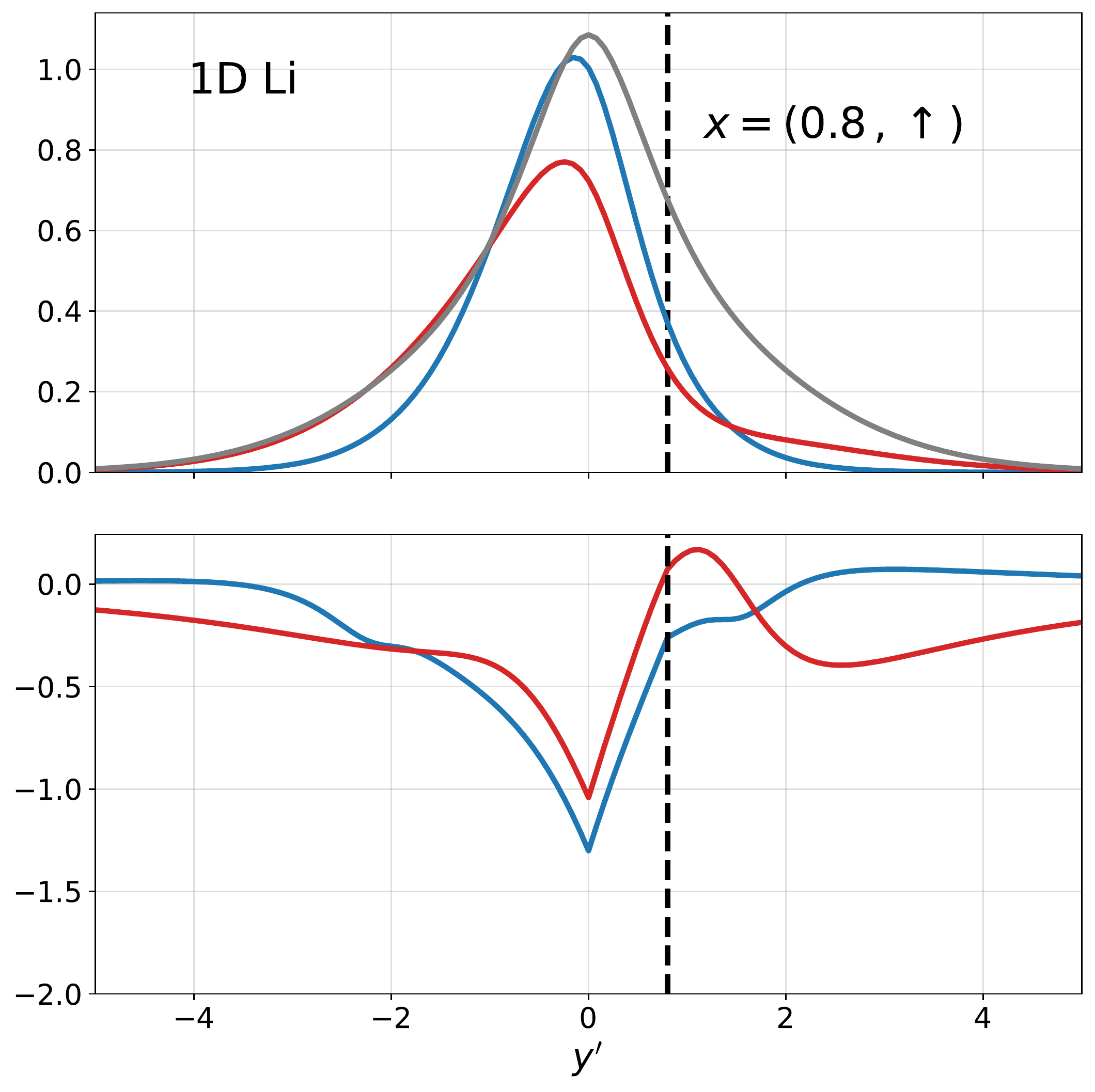}} 
& \subfloat{\includegraphics[width=0.48\textwidth]{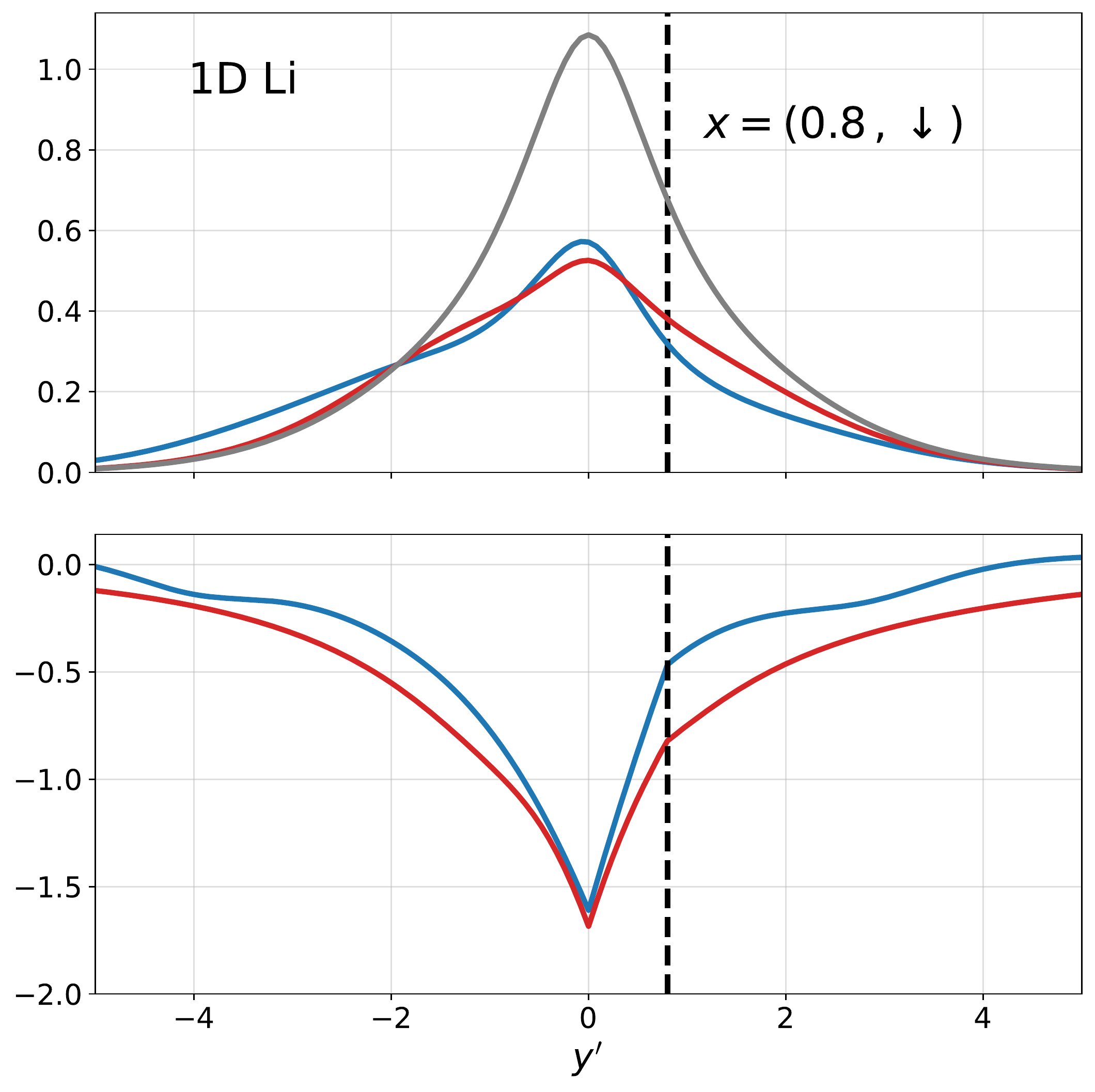}}\\
\end{tabular}
\end{tabularx}

\parbox{\textwidth}{\caption{Spin-CP densities and potentials in 1D Li with $N_{\uparrow} = 2, N_{\downarrow} = 1$, and $\lambda = 1$: $\tilde v_{{\s}, x}^{\mathrm{BEA, LDA}}(y')$ is the 1D BEA approximation to the spin-CP-KS potential using the 1D LDA~\cite{baker2015one} approximation, see Eq.~\eqref{eq: blue el cp-ks potential w lda}, with corresponding spin-CP density $\tilde v_{{\s}, x}^{\mathrm{BEA, LDA}}(y')$. The exact CP-KS potential is $\tilde v_{{\s}, y}(y')$ with corresponding exact CP density $\tilde n_y(y')$, $v_{\s}(y')$ the exact KS potential with corresponding exact ground-state density $n(y')$, and $v(y')$ the external potential for our 1D Li atom. \label{fig: li example with bea}}}
\end{figure*}

\setlength{\tabcolsep}{12pt}
\begin{table*}[!htp]
\centering
\renewcommand{\arraystretch}{2}
\begin{tabular}{ccccccc}
\hline
\hline
 $N$ &     Symbol & $V_{\ee}[n, \tilde n^{\text{BEA}}]$ & $V_{\ee}[n^{\text{LDA}}, \tilde n_{x}^{\text{BEA, LDA}}]$ & $V^{\text{HF}}_{\ee}$ &  $V^{\text{exact}}_{\ee}$ \\
\hline
     2 &         He &                      0.659 (-0.031) &                                                                        0.650 (-0.040) &         0.722 (0.032) &                     0.690 \\
     2 &     Li$^+$ &                      0.739 (-0.016) &                                                                          0.735 (-0.020) &         0.773 (0.018) &                     0.755 \\
     2 &  Be$^{++}$ &                      0.779 (-0.013) &                                                                          0.778 (-0.014) &         0.802 (0.010) &                     0.792 \\
     3 &         Li &                                   - &                                                                            1.741 (0.094) &         1.682 (0.035) &                     1.647 \\
     3 &     Be$^+$ &                                   - &                                                                           1.971 (0.113) &         1.881 (0.023) &                     1.858 \\
     4 &         Be &                                   - &                                                                          3.314 (0.126) &         3.360 (0.172) &                     3.188 \\
\hline
\hline
\end{tabular}
\parbox{\textwidth}{\caption{
Electron-electron repulsion energies, $V_{\ee}$, for 1D systems using various methods. All energies are given in Hartree units. Errors with the exact are given in parenthesis. For $N=3$ (spin-polarized systems), spin-CP-DFT calculations were performed self-consistently for each spin channel. 1D Hartree-Fock (HF) results are given as reference. Exact results are from 1D DMRG calculations~\cite{baker2015one}. The local BEA of the main text performs better for 3D Coulomb-interacting systems~\cite{mccarty2020electronic}.}\label{tab: bea ions}}
\end{table*}

\begin{figure*}[!ht]
\includegraphics[width=\textwidth]{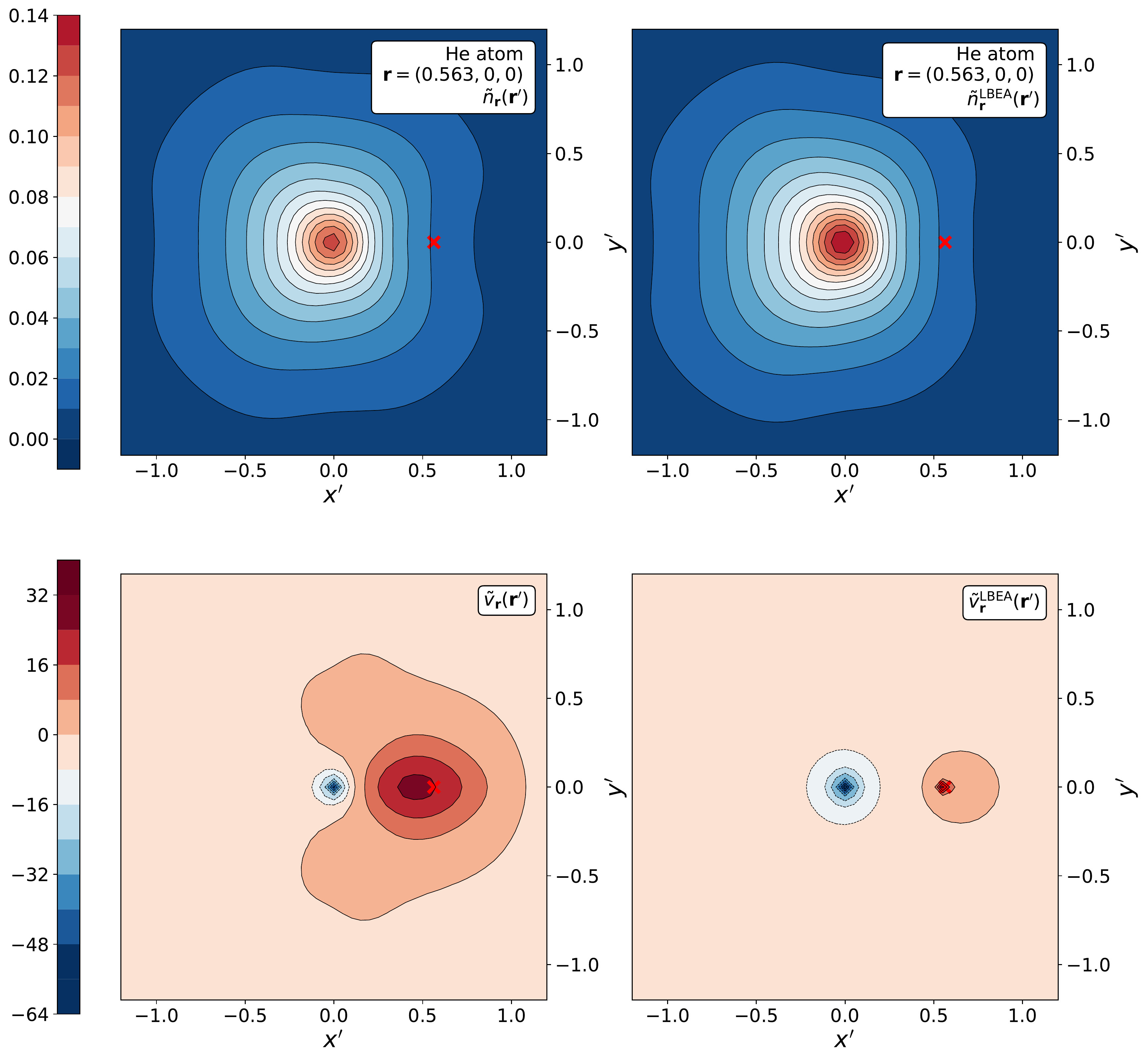}
\centering
\parbox{\textwidth}{\caption{
CP densities and potentials for (3D) He atom ($\lambda = 1$): contour plots for the exact CP density $\tilde n_{\br}(\br')$ (top left) and corresponding exact CP-KS potential $\tilde v_{\br}(\br')$ (bottom left) are plotted within the $(x',y',z'=0)$ plane for reference position $\br = (0.563, 0, 0)$ (displayed as red cross). The corresponding LBEA results are plotted on the right. 
\label{fig: 3D he example with bea}}}
\end{figure*}

The BEA was proposed for 3D Coulomb-repelling electrons, for which it works well~\cite{mccarty2020electronic}. For 1D exponential repulsion, it is less accurate. In 1D the BEA takes the analog, $\Delta \tilde v_{y}^{\lambda, \mathrm{BEA}}(y') = \lambda \, A \exp(-\kappa |y - y'|)$. For $\lambda = 1$ we plot the BEA CP potential and density for 1D He in Figure~\ref{fig: he example with bea}. ~{We see that the BEA is reasonably accurate, but it is clear that the BEA potential, Eq.~\eqref{eq: blue el approx}, is overly repulsive and {overestimates the} CP potential at the reference position.} For $N > 2$ and $\lambda = 1$ we use the 1D LDA approximation (parameterized in ~\cite{baker2015one}) and solve the CP-KS equations self-consistently with the following CP-KS potential:
\begin{equation}
    \tilde v_{{\s}, y}^{\mathrm{BEA}, \mathrm{LDA}}(y') = v(y') + A \exp(-\kappa |y - y'|) + v_{\Hxc}^{\mathrm{LDA}}[\tilde n_{y}](y') \, .
\label{eq: blue el cp-ks potential w lda}
\end{equation}
We plot this potential for 1D Be in Figure~\ref{fig: be example with bea}.~{The BEA results here are worse than in 1D He, {because of} the exchange-limit deficiency previously discussed.} 

We can also apply this potential in the spin-CP-DFT case to spin-polarized systems, such as 1D Li in Figure~\ref{fig: li example with bea}. ~{For $\sigma = \downarrow$, the BEA spin-CP density closely approximates the exact and the BEA does yield an accurate exchange-limit spin-CP density, $n_{\uparrow}^{\text{LDA}}{(y')}$, which is similar to the exact, $\tilde n_{(y, \downarrow)}^{\lambda = 0}{(y')}~=~n_{\uparrow}{(y')}$.}

With $\lambda = 1$ only we can calculate the total electron-electron repulsion energy, $V_{\ee}$, given the total ground-state density $n$ and CP density $\tilde n_{y}$. In our 1D analog we have:
\begin{equation}
    V_{\ee}[n, \tilde n_{y}] = \half \int dy \, n(y) \int dy' \, \tilde n_y(y') \, A \exp(-\kappa |y - y'|) \, .
\end{equation}
In {Table~\ref{tab: bea ions}} we tabulate total electron-electron repulsion energies obtained from various methods. While we note that the BEA is an exceptionally crude approximation in 1D, for two-electron ions it yields energies that have comparable errors to Hartree-Fock (HF).  

In 3D reality with Coulombic interactions, it can be seen from Eq.~\eqref{eq: v cp cusp} that the BEA yields a CP density with a cusp that is too large by a factor of 2. To remedy this, we interpolate with a local density approximation 
\begin{equation}
    \Delta \tilde v^{\lambda, \mathrm{LBEA}}_{\br}[\n] (\br')  \equiv \frac{\lambda}{2|\br-\br'|}
    \left(1+{\rm Erf}\left(\frac{|\br-\br'|}{r_s(\n(\br))}\right)\right),
\label{eq: interpolated bea}
\end{equation}
where $r_s=(3/(4\pi\n))^{1/3}$ is the Wigner-Seitz radius at the reference point $\br$. This \emph{local blue electron approximation} (LBEA) will yield the exact cusp condition by Eq.~\eqref{eq: v cp cusp} and satisfies the same discussed exact conditions as the BEA. In Figure~\ref{fig: 3D he example with bea} we plot the CP densities and potentials for the (3D) He atom in the $(x', y', z' = 0)$ plane for reference position $\br = (0.563, 0, 0)$. We see that despite noticeable errors in the LBEA CP potential, the corresponding CP density is rather accurate. The exact results were obtained using exact diagonalizations in the recently developed \textit{gausslet} basis set~\cite{qiu2021hybrid,mccarty2020electronic}. In Ref.~\cite{mccarty2020electronic}, using the LBEA in Eq.~\eqref{eq: interpolated bea} with HF-calculated total densities, it was found that $E^{\mathrm{LBEA}}_{\xc} = -1.0736$ Ha (error of $0.007$ Ha from the exact XC energy). See Ref.~\cite{mccarty2020electronic} for results on other two-electron ions, Hooke's atom, and H$_2$ dissociation. Within APDFT, the Overhauser model approximation, which represents a radial screened Coulomb interaction, is a close analog to the LBEA in CP-DFT. In fact, using the Overhauser model, the correlation energy results for two-electron ions in Ref.~\cite{gori2005simple} are quite similar to the LBEA results of Ref.~\cite{mccarty2020electronic}. However, in non-radially symmetric systems or systems with more than two electrons, the Overhauser model and LBEA can produce quite different results. For example, in stretched H$_2$ the Overhauser model produces noticeable energy errors~\cite{gori2009range} whereas LBEA obtains the correct dissociation limit~\cite{mccarty2020electronic}.

The CP-DFT approach can be used to address the uniform electron gas (UEG). Because the UEG is translationally invariant, the dependence on the reference position can be dropped. In Ref.~\cite{mccarty2020electronic} the LBEA with an added repulsive Gaussian potential term was used in CP-DFT calculations to accurately approximate the XC energy per particle of the UEG at all $r_s$ values. In addition, the accuracy does not deteriorate as the temperature of the UEG is raised or if a more primitive method (Thomas-Fermi, as opposed to KS-DFT) is used to calculate CP densities. At zero temperature, the added Gaussian potential is needed at high densities where exchange dominates but is not captured well with the LBEA alone. However, it was also necessary to dampen this Gaussian as $r_s$ is increased, yielding a somewhat empirical procedure but high accuracy results~\cite{mccarty2020electronic}. More recently, Ref.~\cite{perchak2022correlation} considers an alternative approach where a portion of the CP density is fixed. Here the parallel component of the spin-CP density, $\tilde\n_{\sigma}(\br', \sigma)$ in Eq.~\eqref{eq: spin-cp density}, is fixed to the exact exchange limit expression for the UEG (which is known analytically~\cite{D30}). A spin-CP-DFT calculation is then performed for the antiparallel component. As highlighted in Section~\ref{sec: Spin-adapted CP-DFT}, using spin densities in spin-CP-DFT is complicated, as these components do not integrate up to integer particle numbers in general{, see Eq.~\eqref{eq: integrate fully-spin decomposed densities}.} However, in the setting of the UEG this is not a concern, as the particle number is infinite (or exceedingly large, in a practical calculation). Furthermore, the parallel component of the spin-CP density is fixed to the exchange limit, which does yield an integer particle number. By construction, this approach recovers the correct exchange energy in the high-density limit, and yields sensible accuracy for all other $r_s$ in the UEG~\cite{perchak2022correlation}.

\clearpage

\sec{Conclusions}
\label{sec: conclusions}
In this work we present CP-DFT as a formally exact theory. CP-DFT is an approach to directly calculate CP densities $\tilde n_{\br}(\br')$ from ground-state calculations of $N - 1$ electrons in an external potential $\tilde v_{\br}(\br')$, known as the CP potential. In practice, the CP potential (or CP correction potential defined in Eq.~\eqref{eq: cp correction potential}) used to obtain the CP density must be approximated. The XC energy can be extracted directly from the adiabatic connection formula using the ground-state density and CP densities. Notably, the CP-DFT approach bypasses the need for an XC energy functional approximation: only accurate ground-state densities and CP densities are needed to yield accurate XC energies within CP-DFT. We also introduce a formally exact spin-CP-DFT approach which can be used to obtain spin-CP densities. Similar to standard spin-DFT, we anticipate that the spin-CP-DFT approach may be more amenable to approximations that can address spin-polarized systems with more accuracy.

Throughout we present several exact conditions associated with CP densities and corresponding CP potentials, including: explicit expressions for two electron systems, long- and short-range asymptotic limits and conditions for general systems, the strictly correlated electron limit, and arbitrary {neutral} hydrogen chains at the dissociation limit.

We highlight an approximation, the ~\emph{blue electron approximation} (BEA), which has semiclassical origins but, surprisingly, satisfies many exact conditions presented. An interpolated variation of BEA, the \emph{local blue electron approximation} (LBEA), additionally satisfies the electron-electron cusp condition in resulting CP densities. For illustrative purposes, we provide select results for the BEA applied on 1D model systems for He, Li, and Be, as well as the LBEA applied on the (3D) He atom. {Because the BEA was designed for 3D Coulomb-repelling systems, it works less well for 1D exponential repulsions, so we strongly recommend against using the 1D-mimic to study the efficacy of BEA.} In previous work~\cite{mccarty2020electronic}, the LBEA was shown to yield usefully accurate results in several systems where standard DFT approximations can fail, such as single electron systems, stretched H$_2$, and the hydrogen anion. In developing generalizeable CP-DFT approximations, it is clear that additional corrections are needed to obtain accurate CP and spin-CP densities in the exchange limit, which is not well captured using the BEA alone, except in two-electron singlet systems. 

Exact conditions might play an important role in guiding construction or testing approximations in CP-DFT. In standard DFT, exact conditions on the energy functionals are most useful, rather than conditions on the densities or potentials.  However, in CP-DFT, where an accurate CP density is the sole interest, only exact constraints on the CP densities and CP potentials are useful. Identifying further exact conditions on the CP density (or equivalently, the XC hole) and associating them to conditions on the CP potential may be helpful in the development of future CP-DFT approximations.

Overall, this work aims to provide a sound theoretical basis for CP-DFT and spin-CP-DFT. Simple examples are used throughout to demonstrate concepts and facilitate understanding. We hope that the content of this work will prompt future approximations and applications.

\sec{Acknowledgments}
Work supported by DOE DE-SC0008696.

\clearpage

\bibliographystyle{apsrev4-2}
\bibliography{Master}

\label{page:end}
\end{document}

%% file: Burke_template.tex
\usepackage{amsmath}
\usepackage{physics}
\usepackage{amssymb}
\usepackage{hyperref}
\usepackage{float}
\usepackage[dvipsnames]{xcolor}

\usepackage[caption=false]{subfig}
\usepackage{tabularx}

\definecolor{TITLECOL}{rgb}{0.05,0.25,0.85}
\definecolor{CONTENTSCOL}{rgb}{0.1,0.2,0.7}
\definecolor{URLCOL}{rgb}{0,0.52,0.83}
\definecolor{LINKCOL}{rgb}{0.05,0.5,0}
\definecolor{CITECOL}{rgb}{0.25,0,0.48}
\definecolor{SECOL}{rgb}{0.07,0.31,0.80}
\definecolor{SSECOL}{rgb}{0.26,0.19,0.75}

\newcommand{\coloredtitle}[1]{\title{\textcolor{TITLECOL}{#1}}}
\newcommand{\coloredauthor}[1]{\author{\textcolor{CITECOL}{#1}}} 
\renewcommand{\sec}[1]{\section{\textcolor{SECOL}{#1}}}

\usepackage{graphicx}

\def\PublicationsLink{http://dft.uci.edu/publications.php}
\def\preprintlink{ \href{\preprintlinklocation}{\TitleOfPaper} }
\def\preprinttext{~}

\usepackage{fancyhdr}
\makeatletter
\fancypagestyle{titlepage}
{	\lhead{\textsc{\preprinttext\ ~ \href{\PublicationsLink}{~}}}
	\chead{}
	\rhead{\preprintlink}
	\lfoot{}}
\makeatother
\def\preprintlink{ 
	\href{\preprintlinklocation}
        {
~}
	}

%% file: KieronsMacros.tex
\def\n{n}

\def\sss{\scriptscriptstyle\rm}

\def\l{^\lambda}

\def\1var{(\bx_1...\bx\N)}

\def\half{\frac{1}{2}}

\def\br{{\bf r}}

\def\bu{{\bf u}}
\def\bx{{x}}

\def\c{\sss C}
\def\s{\sss S}
\def\xc{\sss XC}
\def\Hx{\sss HX}
\def\Hxc{\sss HXC}

\def\N{_{\sss N}}
\def\H{_{\sss H}}

\def\sce{{\rm SCE}}

\def\ee{{\rm ee}}

\def\sph_int{ {\int d^3 r}}

\def\intr{\int d^3r\,}
\def\intrp{\int d^3r'\,}

\def\ncpr{\tilde\n_{\br}}
\def\ncprl{\ncpr^\lambda}

\def\cpr{_{\br}}
\def\cprl{\cpr^\lambda}